\documentclass[sn-mathphys]{sn-jnl}

\usepackage{verbatim}

\jyear{2023}%

\theoremstyle{thmstyleone}%

\theoremstyle{thmstyletwo}%

\theoremstyle{thmstylethree}%

\begin{document}

\title{Constructing phase diagrams for defects by correlated atomic-scale characterization}

\author[1]{\fnm{Xuyang} \sur{Zhou}}\email{x.zhou@mpie.de}
\equalcont{These authors contributed equally to this work.}

\author[1]{\fnm{Prince} \sur{Mathews}}\email{mathews@mpie.de}
\equalcont{These authors contributed equally to this work.}

\author[2]{\fnm{Benjamin} \sur{Berkels}}\email{berkels@aices.rwth-aachen.de}

\author[1]{\fnm{Saba} \sur{Ahmad}}\email{saba@mpie.de}

\author[2]{\fnm{Amel Shamseldeen Ali} \sur{Alhassan}}\email{Alhassan@aices.rwth-aachen.de}

\author[3]{\fnm{Philipp} \sur{Keuter}}\email{keuter@mch.rwth-aachen.de}

\author[3]{\fnm{Jochen M.} \sur{Schneider}}\email{schneider@mch.rwth-aachen.de}

\author[1]{\fnm{Dierk} \sur{Raabe}}\email{raabe@mpie.de}

\author[1]{\fnm{Jörg} \sur{Neugebauer}}\email{neugebauer@mpie.de}

\author[1]{\fnm{Gerhard} \sur{Dehm}}\email{dehm@mpie.de}

\author*[1,4]{\fnm{Tilmann} \sur{Hickel}}\email{hickel@mpie.de}

\author*[1]{\fnm{Christina} \sur{Scheu}}\email{scheu@mpie.de}

\author*[1]{\fnm{Siyuan} \sur{Zhang}}\email{siyuan.zhang@mpie.de}

\affil[1]{\orgname{Max-Planck-Institut für Eisenforschung}, \orgaddress{\street{Max-Planck-Straße 1}, \city{Düsseldorf}, \postcode{40237}, \country{Germany}}}

\affil[2]{\orgdiv{Aachen Institute for Advanced Study in Computational Engineering Science (AICES)}, \orgname{RWTH Aachen University}, \orgaddress{\street{Schinkelstraße 2}, \city{Aachen}, \postcode{52062}, \country{Germany}}}

\affil[3]{\orgdiv{Materials Chemistry}, \orgname{RWTH Aachen University}, \orgaddress{\street{Kopernikusstraße 10}, \city{Aachen}, \postcode{52074}, \country{Germany}}}

\affil[4]{\orgname{Federal Institute for Materials Research and Testing (BAM)}, \orgaddress{\street{Richard-Willstätter-Straße 11}, \city{Berlin}, \postcode{12489}, \country{Germany}}}

\abstract{
Phase transformations and crystallographic defects are two essential tools to drive innovations in materials.
Bulk materials design via tuning chemical compositions has been systematized using phase diagrams. 
We show here that the same thermodynamic concept can be applied to understand the chemistry at defects.
We present a combined experimental and modelling approach to scope and build phase diagrams for defects.
The discovery was enabled by triggering phase transformations of individual defects through local alloying, and sequentially imaging the structural and chemical changes using atomic-resolution scanning transmission electron microscopy.
By observing atomic-scale phase transformations of a Mg grain boundary through Ga alloying, we exemplified the method to construct a grain boundary phase diagram using ab initio simulations and thermodynamic principles.
The methodology enables a systematic development of defect phase diagrams to propel a new paradigm for materials design utilizing chemical complexity and phase transformations at defects.
}


\maketitle

\section{Introduction}\label{intro}

Materials development is the backbone of advancement of human civilization. 
Mastering multi-phase materials and their associated transformations allows for customized applications in infrastructure, transportation \cite{Tasan2015}, energy \cite{Sharma2009}, and medical devices \cite{Raabe2007}.
As a well-organized, systematic thermodynamic approach to design materials, phase diagrams serve as a crucial instrument for comprehending the effects of variables such as temperature, pressure, and chemical composition on phases and properties \cite{Gibbs1948}. 

While phase diagrams predict the constituent phases and their coexistence in materials, they do not account for crystallographic defects which however control many materials properties. 
Examples are dislocations and grain boundaries (GB) which determine many mechanical \cite{Buban2006,Lu2000}, transport \cite{Legros2008} and functional properties \cite{Lu2004}.
In addition to the local structural distortions that are characteristic of defects, their chemical composition can also deviate significantly from that of the surrounding bulk phase. 
For example, solute elements can segregate around dislocations (Cottrell atmospheres \cite{Cottrell1949,Kuzmina2015,Zhou2021,Yu2022}), stacking faults (Suzuki effect \cite{Suzuki1962,Palanisamy2019}), and GBs \cite{Lejcek1991,Kirchheim2002,Wang2011,Schuh2012,Nie2013,Raabe2014,Yu2017,Lejcek2017,Zhou2023}.
Such effects have been increasingly employed, as materials design goes through a paradigm shift to exploit rather than avoid the chemical complexity around defects. 
Such confined and low-dimensional chemical states have been earlier termed as ``low-dimensional phases'' \cite{Hart1968,Frolov2015,Brink2022}, ``complexions'' \cite{Dillon2007,Harmer2011,Luo2011,Kaplan2013,Cantwell2014,Yu2017}, or ``defect phases'' \cite{Korte-Kerzel2022} to differentiate them from bulk phases.

GBs separate regions of individual single-crystalline grains and can control functional properties such as electrical resistivity \cite{Lu2004,Bishara2021,BuenoVilloro2023,BuenoVilloro2023a}, magnetic coercivity \cite{Duerrschnabel2017}, as well as mechanical strength and ductility \cite{Hall1951,Wu1994,Buban2006,Luo2011,Khalajhedayati2016,Krause2018,Cantwell2020,Dehm2022}.
The thermodynamic theory on GB phase transformations \cite{Hart1968,Cahn1982,Rottman1991,Tang2006} has been developed further in the last decade by Frolov and Mishin \cite{Frolov2012,Frolov2012a}. 
It has been recently revealed that different defect phases can coexist at GBs even in elemental metals with only a single bulk phase, such as Cu \cite{Frolov2013,Meiners2020,Frommeyer2022}.
Moreover, introducing alloying elements can be utilized to control these defect phases, their transformations and in turn the properties, opening new doors to evolve materials design \cite{Schuh2012,Cantwell2020}.
Different defect systems have demonstrated phase transformations. 
For example, on the mesoscopic scale, GB faceting transformation was observed in Cu GBs, triggered by adding Bi \cite{Ference1988} or Ag \cite{Peter2021}.  
In another case, segregation of alloying elements or impurities can lead to embrittlement of the metal through the formation of a liquid GB phase \cite{Luo2011,Sigle2006,Lejcek2017,Zhao2022}.

A systematic exploration and design of defect phases requires guidance by thermodynamic defect phase diagrams \cite{Korte-Kerzel2022}. 
Phase diagrams for defect-free bulk materials are commonly developed through the synthesis and characterization of samples with varying compositions.
However, this methodology does not work for defect phases, because for a single bulk composition, the elements partition differently among the bulk phase and various types of defects.
As the elemental partition reaches local thermodynamic equilibrium, the chemical potential becomes constant across the bulk and defect phases, which can be used as the universal state variables to construct defect phase diagrams \cite{Korte-Kerzel2022}.

For an experimental construction of defect phase diagram, the same defect must be examined at adjustable chemical potentials.
We present here a local alloying approach to tune the chemical potentials at a GB. 
By atomic-scale scanning transmission electron microscopy (STEM) imaging, we have identified transformations of a single GB between different structures as well as various types of ordering of the alloy element. 
Density functional theory (DFT) calculations were then performed to evaluate the thermodynamic stability of the observed GB phases, which led to the construction of a defect phase diagram for the GB.
The methodology can be employed to populate defect phase diagrams for many more nanoscopic defects crucial to the design of modern materials.

\section{Results and Discussion}\label{res}
\subsection{GB phase transformations triggered by local alloying}

\begin{figure}[h]%
\centering
\includegraphics[width=0.98\textwidth]{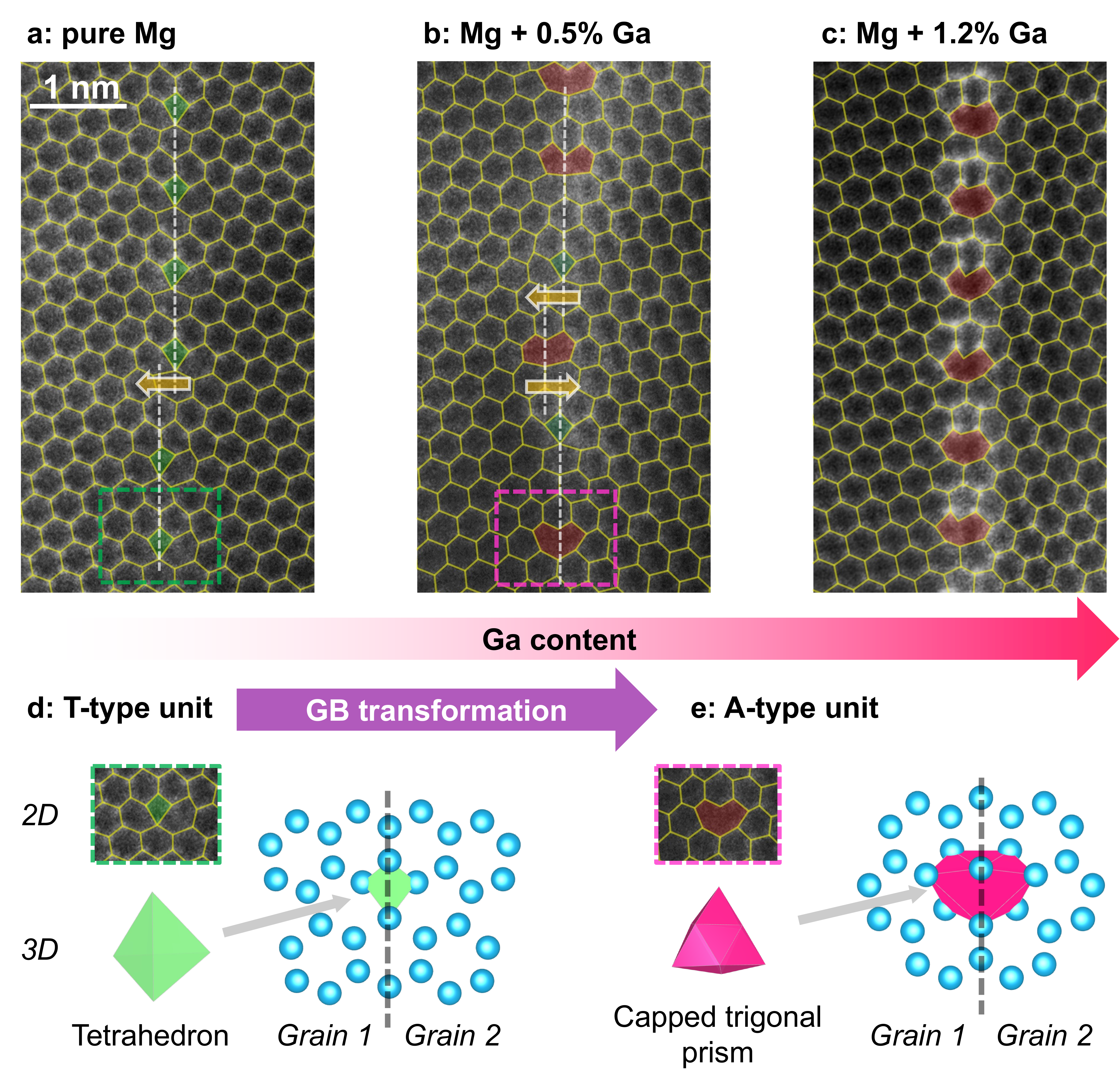}
\caption {\textbf{Experimental observation of a GB phase transformation in Mg by local alloying of Ga.} (a-c) High angle annular dark field (HAADF)-STEM images of a $\Sigma$7 GB in pure Mg, (b) Mg-0.5at.\%Ga and (c) Mg-1.2at.\%Ga. The images are overlaid with polygon grids to highlight the GB structural units (Section~\ref{sec:PatternRec}), while raw experimental images are presented in Fig.~\ref{transformStructure_SI}. The dashed lines in (a) and (b) highlight the GB planes and the arrows point to shifts in the GB plane. (d) Atomistic structures of the T-type structural unit (green tetrahedron) and (e) the A-type structural unit (pink capped trigonal prism) in 2D and 3D.}\label{transformStructure}
\end{figure}

To study GB phase transformation on the atomic scale, we selected symmetric GBs in hexagonal close-packed (HCP) Mg along the [0001] tilt axis, which can adopt to a variety of atomistic structures. 
In particular, the $\Sigma$7 [0001]${\{3\overline{1}\overline{2}0\}}$ GBs (abbreviated as $\Sigma$7 GBs) have a misorientation angle of $\approx$22° \cite{Zhang2022}, and two types of structural units, T-type and A-type \cite{Wang1997}.
Experimentally, we have identified one $\Sigma$7 GB in HCP Mg based on 4D-STEM mapping (see Fig.~\ref{4dSTEM}).
As shown in Fig.~\ref{transformStructure}a, the $\Sigma$7 GB is atomically resolved by STEM imaging, where T-type structural units are found along the GB plane.
Such T-type units are common structural units to build GBs \cite{Ashby1978,Pond1979,Sutton1989}, which have a tetrahedron shape in 3D, as shown in Fig.~\ref{transformStructure}d.
For $\Sigma$7 GBs, each structural unit is periodically spaced at 0.85~nm and can be viewed as a dislocation core with a Burgers vector of $\frac{1}{3}\langle2\overline{1}\overline{1}0\rangle$ (Fig.~\ref{loop}a). 
Beside the T-type units, GB disconnections (e.g., arrow in Fig.~\ref{transformStructure}a) are also found along the GB to account for the shift of GB plane. 

We then applied local alloying by Ga\textsuperscript{+} implantation and subsequent diffusion to the Mg sample and returned to the same $\Sigma$7 GB for high resolution imaging (Details in Section~\ref{sec:implantation} and Fig.~\ref{implantation}). 
As shown in Fig.~\ref{transformStructure_SI}b and c, the GB areas show a higher atomic number (Z) contrast than the surrounding grains, demonstrating the tendency of Ga atoms to segregate at GBs \cite{Zhang2022}. 
Closer examination reveals that after local alloying of 0.5at.\% Ga, some GB structural units have been transformed to A-type units (Fig.~\ref{transformStructure}b), and after local alloying of 1.2at.\% Ga, the GB has been fully transformed to A-type units (Fig.~\ref{transformStructure}c). 
The A-type unit has a larger core structure and a capped trigonal prism shape in 3D (Fig.~\ref{transformStructure}e), where the same Burgers vector $\frac{1}{3}\langle2\overline{1}\overline{1}0\rangle$ can be constructed (Fig.~\ref{loop}b).

In pure Mg, the formation energies of $\Sigma$7 GBs composed of T-type and A-type structural units are evaluated by DFT calculations as 0.309 and 0.311~J~m\textsuperscript{-2}, respectively. 
The experimentally observed T-type unit is indeed the ground state configuration for Mg $\Sigma$7 GB, although the formation energy for A-type $\Sigma$7 GB is only marginally higher, as was reported in previous DFT calculations on Mg \cite{huber_Mg} and ZnO \cite{Sato2007} $\Sigma$7 GBs.
To understand the structural transformation of the GB, we simulated the local alloying by replacing one Mg column in the T-type unit by Ga and studying their structural relaxation by DFT. 
The atomic sites for the structural units are labelled in alphabetical order with respect to the increasing distance to the GB plane, as shown in Fig.~\ref{loop}a, b.
The T-type unit with the \textbf{b2} site occupied by Ga atoms spontaneously relaxes into an A-type unit with Ga atoms occupying the \textbf{a1} site, as can be seen in Fig.~\ref{TtoA_DFT}a. 
A similar structural transformation sequence with Ga occupying a neighboring site is presented in Fig.~\ref{TtoA_DFT}b. 
In each relaxation step, the identical DFT structure is viewed from two perspectives, matching to the T-type (filled green shape) and A-type (open pink shape) structural units. 
As the GB structure relaxes, the distortion of the T-type unit increases, while the A-type unit becomes more symmetric. 
This visualization shows the shift of the local symmetric structure from the T-type to the A-type structural unit, completing the GB phase transformation. 

To trace the GB phase transformation from STEM images, we have developed automatic pattern recognition to classify experimental GB structural units into T-type and A-type, as overlaid in Fig.~\ref{transformStructure}a-c.
As schematically shown in Fig.~\ref{TtoA_Pattern} and detailed in Section~\ref{sec:PatternRec}, DFT structures serve as inputs to locate GB structural units in the STEM images. 
Then the positions of the atomic columns are labelled and compared with the DFT structures to reach a decision on a better match to T-type or A-type units.
Moreover, it is observed from Fig.~\ref{TtoA_DFT} that after the structural transformation, the GB plane is shifted by $\frac{1}{6}\langle2\overline{1}\overline{1}0\rangle$, half of the Burgers vector. 
This is experimentally captured between the middle A unit and its adjacent T units, as pointed by arrows in Fig.~\ref{transformStructure}b. 

\subsection{GB phase transformation of different chemical ordering}

\begin{figure}[h]%
\centering
\includegraphics[width=0.98\textwidth]{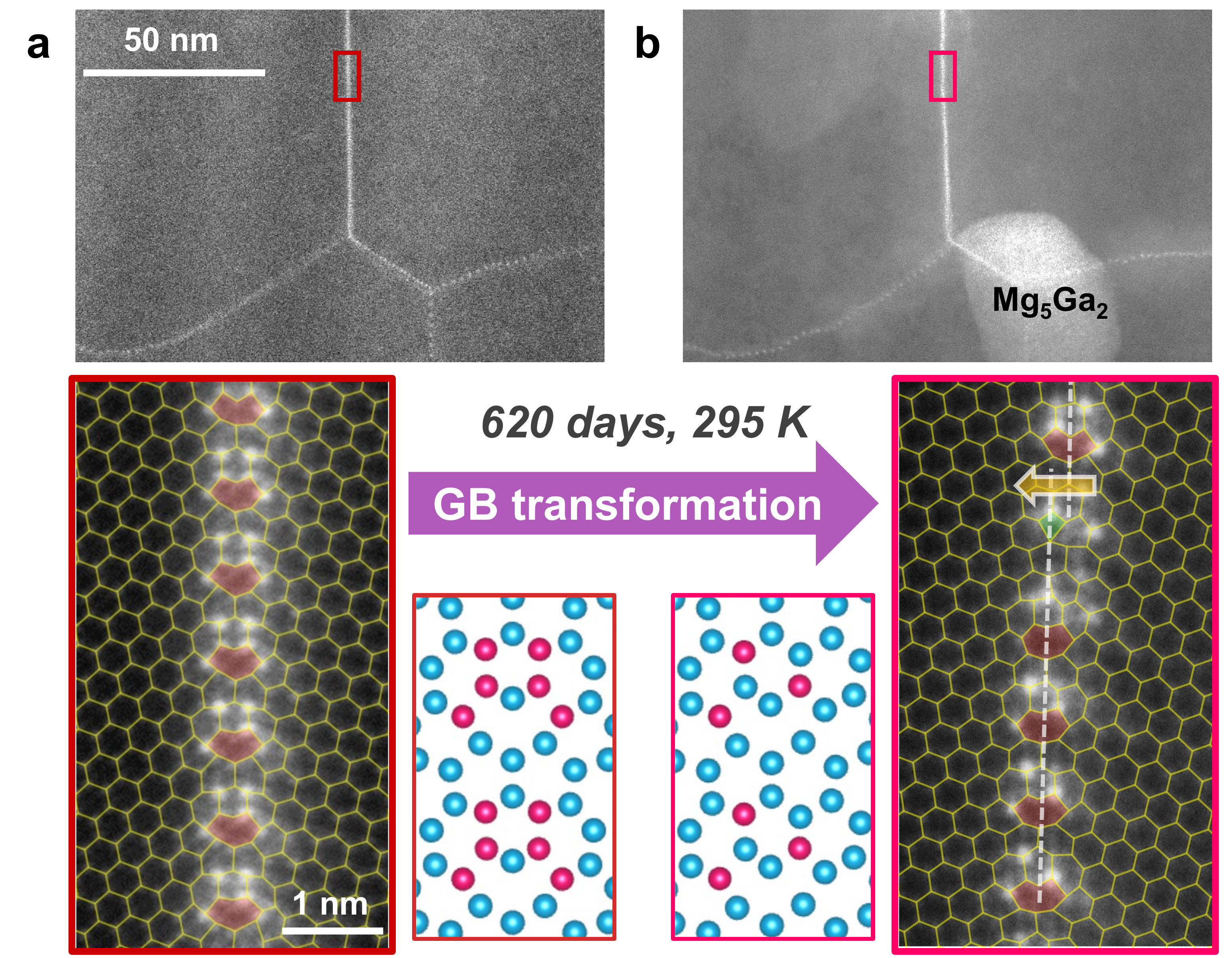}
\caption{\textbf{Transformation of chemically-ordered GB phases.} HAADF-STEM images of the same $\Sigma$7 GB (a) 1 day and (b) 620 days after Ga\textsuperscript{+} beam thinning. The long storage time enabled excess Ga atoms on the surface to form bulk Mg\textsubscript{5}Ga\textsubscript{2} precipitates, while the ordered 6-Ga unit (a) transforms to a differently ordered 3-Ga unit (b). The Ga atoms in the atomic configurations are highlighted in pink, with their atomic sites labelled. The A-type structural units are highlighted by pink capped trigonal prisms. The corresponding STEM images without overlaid grids are presented in Fig.~\ref{transformTime_SI}. }\label{transformTime}
\end{figure}

With increasing amount of local alloying, chemical ordering of Ga atoms start to appear, as shown by the brighter atomic columns at the GB (Fig.~\ref{transformStructure_SI}c).
The chemical ordering is also reproduced in a TEM sample processed and thinned down completely with focused ion beam (FIB) of Ga\textsuperscript{+}, as an attempt to introduce maximum amount of Ga to the sample, and hence the GBs. 
Ga segregation to the GBs are evidently shown in Fig.~\ref{transformTime}a, and it is clear that the Ga-decorated $\Sigma$7 GB is composed of A-type structural units.
Moreover, there is a clear pattern of Ga segregation to pairs of \textbf{b}, \textbf{c}, and \textbf{e} sites, in total six atomic columns for each structural unit.

After 620 days storage in a desiccator, the same GB underwent a transformation to another ordered phase, as shown in Fig.~\ref{transformTime}b. 
Instead of six, only three bright Ga columns remained in each structural unit. 
The Ga atoms stay on the \textbf{b}, \textbf{c}, and \textbf{e} sites, except that they are no longer present in pairs. 
In the upper part of high resolution image in Fig.~\ref{transformTime}b, the \textbf{b1} site is occupied by Ga along with the \textbf{c2} and \textbf{e2} sites on the right. 
Likewise, the lower part of the image shows the mirrored occupation on \textbf{b2}, \textbf{c1} and \textbf{e1} sites. 
Except for a structural unit at the disconnection, all other units remain to be classified as A-type.
Energy dispersive X-ray spectroscopy (EDS) reveals a Ga content of 0.7 at.\% inside the Mg grains, which increases significantly at the $\Sigma$7 GB (Fig.~\ref{transformTime_SI}g,h).
For this GB, the Gibbsian interfacial excess (shaded area in Fig.~\ref{transformTime_SI}h) is 6.95 Ga atoms/nm$^2$, corresponding to a coverage of 3 Ga atoms per structural unit.
As there are exactly three bright columns in each structural unit, the ordered phase is simulated using full Ga occupancy at the corresponding sites.

\subsection{Construction of the GB defect phase diagram}

\begin{figure}[h]%
\centering
\includegraphics[width=0.98\textwidth]{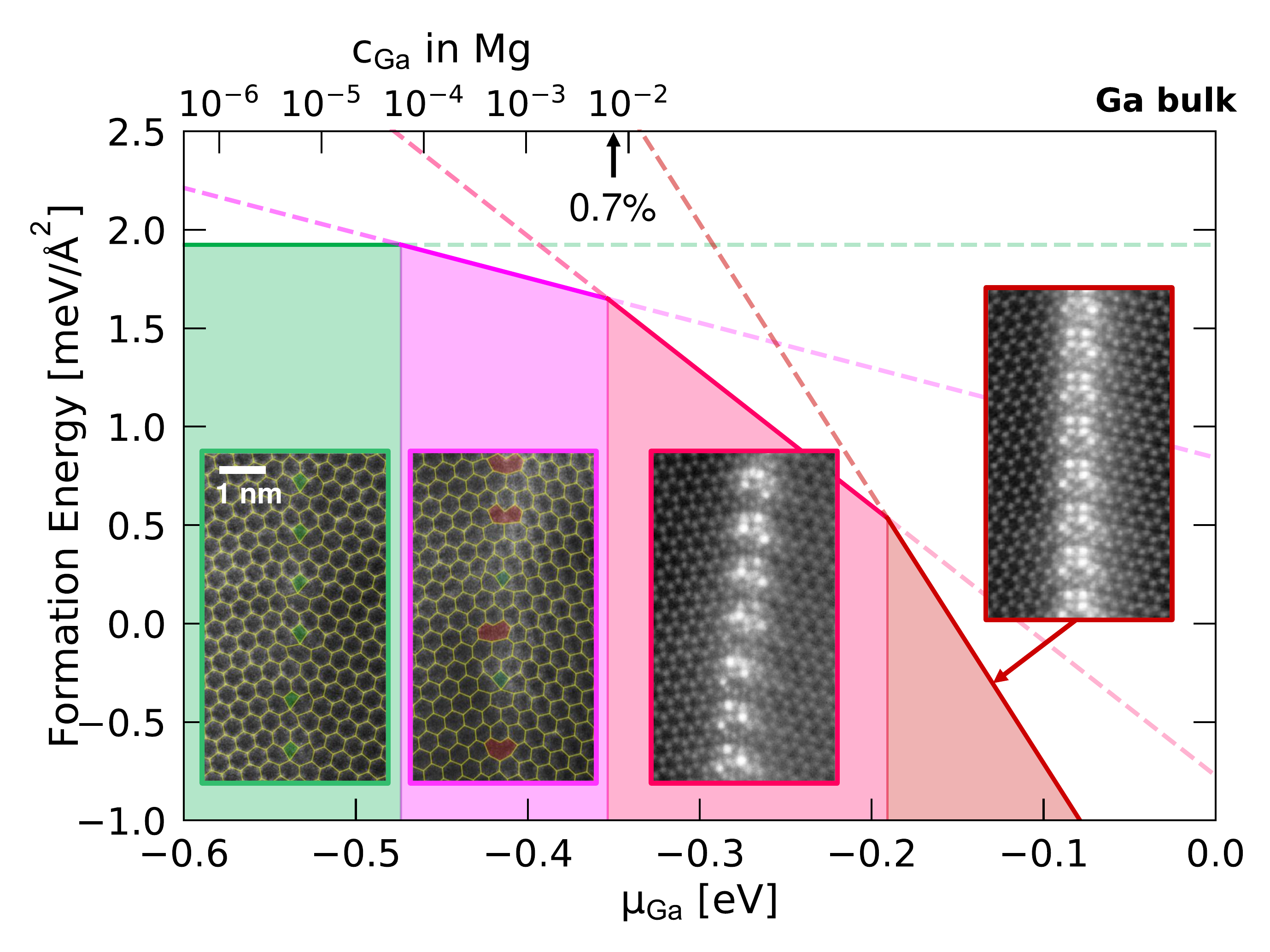}
\caption{\textbf{Construction of a defect phase diagram from observed phase transformations.} The calculated Mg $\Sigma$7 GB phase diagram using DFT and experimentally-observed GB structures. The bottom axis $\mu_{\rm Ga}$ is the chemical potential of Ga with respect to the bulk Ga phase. The top axis provides the corresponding dilute Ga concentration up to 1\% in the Mg solid solution at 300~K. The arrow points to the measured $c_{\rm Ga}=0.7\%$ in the solid solution.
}\label{DPD} 
\end{figure}

Having identified the structural transformation (from T-type to A-type) of Mg $\Sigma$7 GB triggered by local alloying of Ga, as well as the transformation to different ordering of Ga, we investigate the thermodynamic stability of the observed GB phases in the framework of defect phase diagram.
As shown in Fig.~\ref{DPD}, four ordered structures of Mg $\Sigma$7 GBs were simulated by DFT calculations, including respectively 0, 1, 3, and 6 Ga atoms in the supercells.
As the numbers of Mg and Ga atoms differ in each phase, their formation energy is a function of the chemical potentials, $\mu_{\rm Mg}$ and $\mu_{\rm Ga}$ (Eq.~\ref{Eq:Ef}).
As a reference state, $\mu_{\rm Mg}$ and $\mu_{\rm Ga}$ are set to zero when they are in equilibrium with a reservoir of Mg and Ga atoms in their bulk phases, respectively.
As the entire sample is made of Mg, it fulfills the Mg-rich condition $\mu_{\rm Mg}=0$~eV.
On the other hand, the Ga-rich condition $\mu_{\rm Ga}=0$~eV is only achieved when a layer of Ga is formed on the sample surface, e.g. after Ga FIB.

As the formation energy of the four modelled structures is plotted as a function of the chemical potential (Fig.~\ref{DPD}), they are predicted stable in different ranges of $\mu_{\rm Ga}$.
This is because each line of formation energy has a unique slope, which is proportional to the number of Ga atoms, $N_{\rm Ga}$, in the respective GB phases (Eq.~\ref{Eq:Ef}). 
For the GB phase of pure Mg (0-Ga), the absence of Ga makes the formation energy independent of $\mu_{\rm Ga}$, while this phase has the lowest energy at $\mu_{\rm Ga}<-0.473$~eV.
As $N_{\rm Ga}$ increases in the 1-Ga, 3-Ga, and 6-Ga phases, their formation energy decreases more sharply towards the Ga-rich condition, so that they become the most stable phase at $\mu_{\rm Ga}>-0.473$ (1-Ga), $-0.353$ (3-Ga), and $-0.190$~eV (6-Ga), respectively. 

While local alloying spans a range of chemical potentials to explore new GB phases, the modelling of GB defect phase diagram provides the thermodynamic tool to understand defect phase transformations.
As experimental evaluation of chemical potentials is usually not straightforward, we elaborate on the interpretation and design of GB defect phase diagrams.
As shown in Fig.~\ref{DPD}, local alloying of Ga allows surveying the entire range of $\mu_{\rm Ga}$ from the Ga-free end to the Ga-rich conditions.
Due to the limited solubility of Ga in Mg ($<$ 1 \%, all numbers in atomic fractions), excessive Ga can build up a bulk-phase layer on the implanted surface to define $\mu_{\rm Ga}=0$ eV.
At this $\mu_{\rm Ga}$, we have indeed observed the predicted stable GB phase 6-Ga.
However, it is worth noting that during early stages of Ga\textsuperscript{+} implantation and diffusion, there is no thermodynamic equilibrium observed in experiments to simply relate to a $\mu_{\rm Ga}$. 

Therefore, we propose to explore the lower chemical potentials as diffusion enables the transition from local states of thermodynamic equilibrium to more global states.
There are several intermetallic phases in the Mg-Ga phase diagram \cite{Feng2009}, so that direct interfacing Mg with Ga is not the thermodynamic ground state.
After sufficient time, the surface layer of Ga is diffused into Mg matrix to form islands of Mg\textsubscript{5}Ga\textsubscript{2} precipitates (Fig.~\ref{transformTime_SI}b-d), the most Mg-rich intermetallic phase in the Mg-Ga system \cite{Feng2009}. 

To evaluate $\mu_{\rm Ga}$ in this condition, we notice that Ga atoms within the GB is in equilibrium with those in the Mg solid solution.
For dilute Ga within Mg solid solution, the chemical potential can be evaluated by $\mu_{\rm Ga}=-0.222$~eV + $k_BT \ln (c_{\rm Ga})$, where  $-0.222$~eV is the mixing enthalpy of one Ga atom inside the Mg phase, and $k_BT \ln (c_{\rm Ga})$ is the configurational entropy term at Ga concentration $c_{\rm Ga}$.
Hence, $\mu_{\rm Ga}$ is related to the local equilibrium with Ga solutes in Mg, as represented by the top axis of Fig.~\ref{DPD}. 
In particular, at the measured $c_{\rm Ga}=0.7\%$ in the solid solution (marked by arrow in Fig.~\ref{DPD}), the experimentally observed 3-Ga GB phase is indeed predicted to be most stable.
Furthermore, the evaluated transition from 0-Ga (T-type) and 1-Ga (A-type) $\mu_{\rm Ga}=-0.473$~eV corresponds to very low Ga concentration of $c_{\rm Ga}=6\cdot10^{-5}$ at 300~K.
Since this concentration is well within the solubility limit, there is no driving force to further lower $\mu_{\rm Ga}$, and hence the A-type structural units will remain stable for Mg $\Sigma$7 GB. 

\section{Conclusion}\label{conclusion}

The development of phase diagrams that predict the appearance of certain phases and structures as a function of temperature, pressure and chemical composition is one of the biggest success stories in materials science.
It has helped to turn the empirical material design approach into a knowledge-based method that rests on strict thermodynamics.
Here we have translated this principle to another essential tool in material design, namely the lattice defects. 
Specifically, we have demonstrated an effective way to construct defect phase diagrams by combining atomic-scale STEM characterization of defect structures, automatic pattern recognition and DFT modelling of their energetics. 
The chemical potential axis is sampled experimentally by successive steps of local alloying as well as allowing time for diffusion to transit local thermodynamic equilibrium to a more global scale. 
We have driven and monitored the phase transformation of the same Mg $\Sigma$7 GB from a T-type tetrahedron structural unit to A-type capped trigonal prism unit by local alloying of Ga.
Automatic pattern recognition algorithms have been developed to classify GB structural units and enabled tracing the defect phase transformation.
Different GB phases with ordered Ga atoms have been identified using STEM, including a 6-atom configuration in equilibrium with Ga and a 3-atom configuration in equilibrium with Mg-Ga solid solution. 
The experimental exploration covers a full range of chemical potentials relevant for the design of GB phases, and the structural information can be directly fed into atomistic modelling and narrows down the otherwise gargantuan space of configurations. 
DFT calculations not only provide the relative stability of different GB phases, but also connect them to local thermodynamic equilibrium by the chemical potential. 
The developed methodology can be generally applied to study wide ranges of GBs and defects, and expedites the construction of defect phase diagrams for their usage in science and engineering.

\bmhead{Supplementary information}

If your article has accompanying supplementary file/s please state so here. 

\bmhead{Acknowledgments}

This work was supported by the German research foundation (DFG) within the Collaborative Research Centre SFB 1394 “Structural and Chemical Atomic Complexity—From Defect Phase Diagrams to Materials Properties” (Project ID 409476157). 

\begin{appendices}

\section{Extended Data: Methodology}\label{method}

\subsection{Synthesis}
The nanocrystalline Mg thin film was sputter-deposited onto a Si (100) substrate. 
Detailed description on the synthesis conditions are given in \cite{Zhang2022}, as well as the characterization of the sharp basal plane texture of Mg. 
The thin film hence contains numerous Mg [0001] tilt GBs with a random distribution of the misorientation angle \cite{Zhang2022}. 
The $\Sigma7$ GBs were selected by examining GBs with a misorientation of $\approx$22°.

\subsection{Local alloying to trigger defect phase transformation}
\label{sec:implantation}
Focused ion beam (FIB) was applied to introduce the alloying element Ga locally into the area of examined GBs.
First, the Ga-free specimen for STEM was prepared on a plasma FIB (Thermo Fisher) starting with Xe\textsuperscript{+} polishing at 30~kV and ending with a cleaning step a 8~kV. 
The sample has a volume of $6.5 \cdot 3 \cdot 0.2~ {\rm \mu m}^3$, amounting to 280~fmol of Mg atoms. 
A Scios2 FIB (Thermo Fisher) was then employed to introduce Ga\textsuperscript{+} ions into the lamella. 
For that, cleaning cross-section patterns were used with the sample tilted $\pm$8° towards the Ga\textsuperscript{+} beam at 5~kV and 7.7~pA. 
In such condition, the sample receives 0.08~fmol of Ga\textsuperscript{+} ions for each second. 
Suppose an implantation rate of 100\%, 1.7\% of Ga relative to Mg would be introduced to the sample in a minute.
From the measured Ga concentrations in the Mg sample, 0.50\% and 1.24\% after 1~min and 3~min Ga\textsuperscript{+} implantation, respectively, the implantation rates are evaluated as 29\% (after 1~min) and 24\% (after 3~min). 

\subsection{Electron Microscopy}
High resolution STEM imaging was performed on a Titan Themis microscope (Thermo Fisher) operated at 300~kV. 
Using an aberration-corrected STEM probe of less than 0.1~nm size and 23.8~mrad convergence semi-angle, STEM images were acquired using a HAADF detector with collection semi-angles of 62-200~mrad. 
EDS spectrum imaging was acquired using the SuperX detector. 
Multivariate statistical analysis was applied for noise reduction \cite{Zhang2018}, and subsequent elemental quantification was performed using the Cliff-Lorimer method. 
Precession electron diffraction 4D-STEM imaging was performed on a JEM2200 microscope (JEOL) operated at 200~kV. 
The beam size was $\approx$2~nm and a precession angle of 0.5° was used. 

\subsection{Computational Details}\label{DFT_calculations}
DFT calculations in this work have been carried out using the Vienna Ab initio Simulation Package (VASP) \cite{vasp_1,vasp_2} with the projected augmented wave method \cite{vasp_3} to describe the interaction between ionic cores and valence electrons. 
The Perdew-Burke-Ernzerhof form of parameterization of the generalized gradient approximation has been used to describe the exchange-correlation effects. 
A plane wave cutoff energy of 550~eV was used and based on the Monkhorst-Pack scheme \cite{monkhorst_pack}, the Brillouin zone was sampled with a k-point spacing of 0.12~nm\textsuperscript{-1} along all directions for all structures. 
The Methfessel-Paxton \cite{methfessel_paxton} smearing scheme was applied with the smearing width set to 0.15~eV.\par 

After the construction of supercells containing two GBs, they have been optimized by subjecting them to strains in the direction normal to the GB. 
A relaxation of atomic positions is performed in order to get to the equilibrium structure, thus preserving the lattice constants in the bulk regions of the supercell as well. 
The energy for this equilibrium structure is then applied to calculate the GB energy ($\gamma_{\text{GB}}$) using the following equation:
\begin{equation} \label{Eq:GBE}
    \gamma_{\text{GB}} = \frac{E_{\text{GB}} - E_{\text{bulk}}}{2A_{\text{GB}}}, 
\end{equation}
where $E_{\text{GB}}$ is the total energy of the supercell with the GBs, $E_{\text{bulk}}$ is the energy of hcp Mg bulk rescaled according to the number of atoms in the supercells for the GB and $A_{\text{GB}}$ is the area of cross-section of the GB. 
As each GB supercell contains two GBs, a factor of two is included in the denominator.

To analyze the competition of the defect phases with and without Ga addition, we look into the formation energy $E_{\text{f}}$ of the phases that are calculated using the following equation:
\begin{equation} \label{Eq:Ef}
    E_{\text{f}} = \frac{E_{\text{GB}} - N_{\text{Mg}}\mu_{\text{Mg}} - N_{\text{Ga}}\mu_{\text{Ga}}}{2A_{\text{GB}}},
\end{equation}
where $N$ represents the number of Mg and Ga atoms in the supercell, and $\mu$ represents their chemical potentials. 

\subsection{Automatic Pattern Recognition}
\label{sec:PatternRec}
\begin{figure}[h]%
\centering
\includegraphics[width=0.8\textwidth]{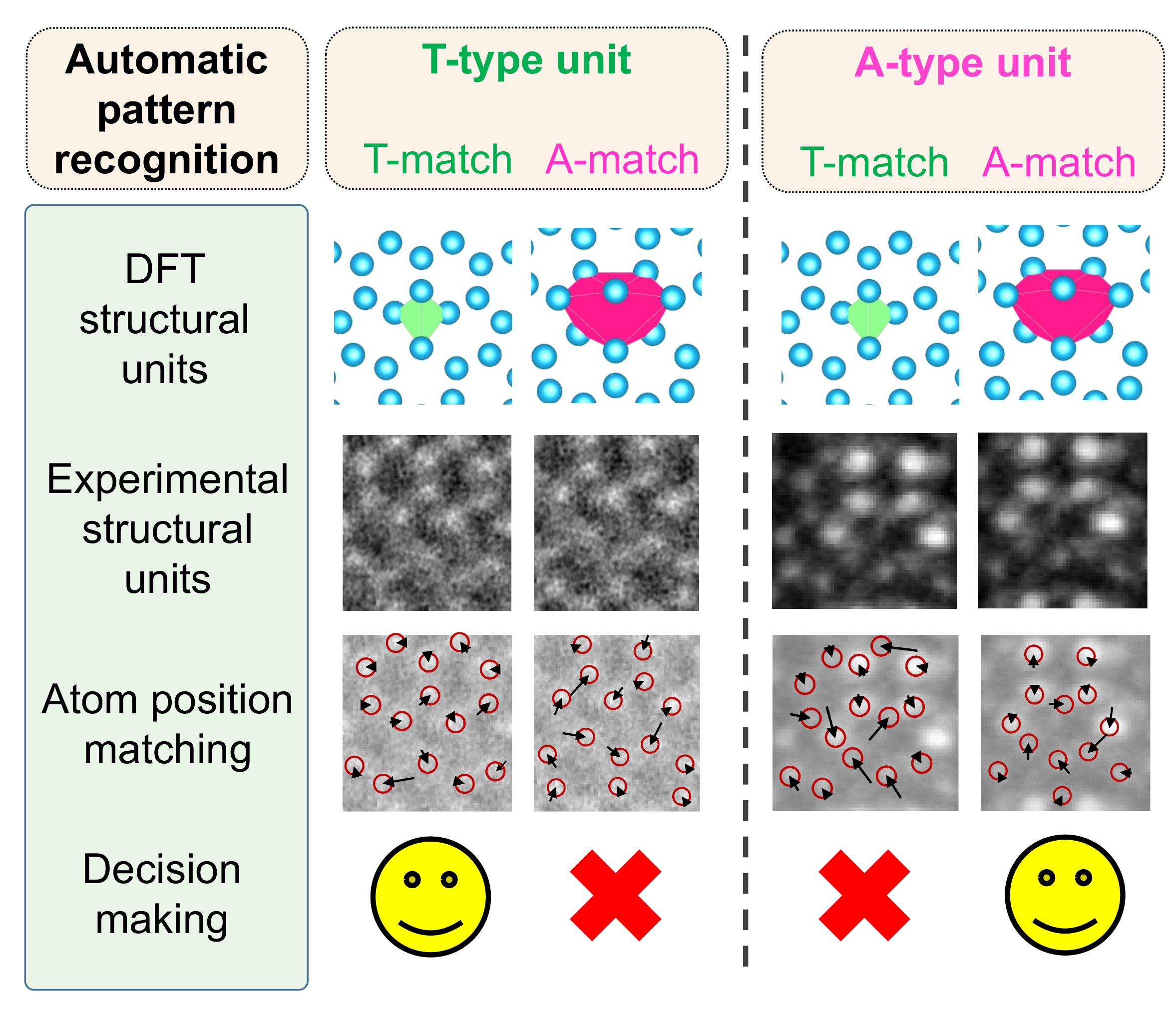}
\caption{\textbf{Automatic pattern recognition to classify experimental images into T-type and A-type structural units.} The displacement vectors between DFT and experimental structures are magnified by five times for visualization.}\label{TtoA_Pattern}
\end{figure}

To detect and distinguish the T-type and A-type structural units in STEM images, we use a novel mathematical framework that detects patterns in the form of atomic arrangements from simulations in atomic scale images and quantifies their differences by estimating the deviation between experimental and computational atomic arrangement.
This framework uses a multi-step procedure. 
First, the simulated atomic arrangement (``DFT structural units'' in Figure~\ref{TtoA_Pattern}) is converted to a synthetic image patch with a sum of Gaussians centered at the simulated positions. 
The potential occurrences of this patch in the STEM image are found with template matching using normalized correlation (see ``Experimental structural units'' for example image patch matches in Figure~\ref{TtoA_Pattern}).
On each matching patch, the deviation between the simulated atomic column arrangement to the positions in the STEM image is estimated using bump fitting (``Atomic position matching'' exemplifies the deviation as shift for each atom in the structural unit in Figure~\ref{TtoA_Pattern}). 
The fitting of the positions from the simulated pattern to the experimental image is split into two steps: First the affine part of the deviation is determined, then the remaining nonlinear part. 

From the resulting fit, we derive feature descriptors that allow to determine which of the given patterns is present. 
Noting that near a candidate location for an A-type match, there are two neighboring candidates for a T-type match. 
We derive the same descriptors on these neighboring positions using the T-type arrangement. 
Only one of the two neighboring matches is reasonable, as the other one can be easily determined from the large deviation necessary to fit the simulation pattern to the image and is discarded (Figure~\ref{TtoA_Pattern} shows only the reasonable match). 
Thus, for each A-type match, we have two sets of descriptors, one describing how well the A-type fits, the other one on the T-type fits. 

\begin{figure}[h]%
\centering
\includegraphics[width=0.8\textwidth]{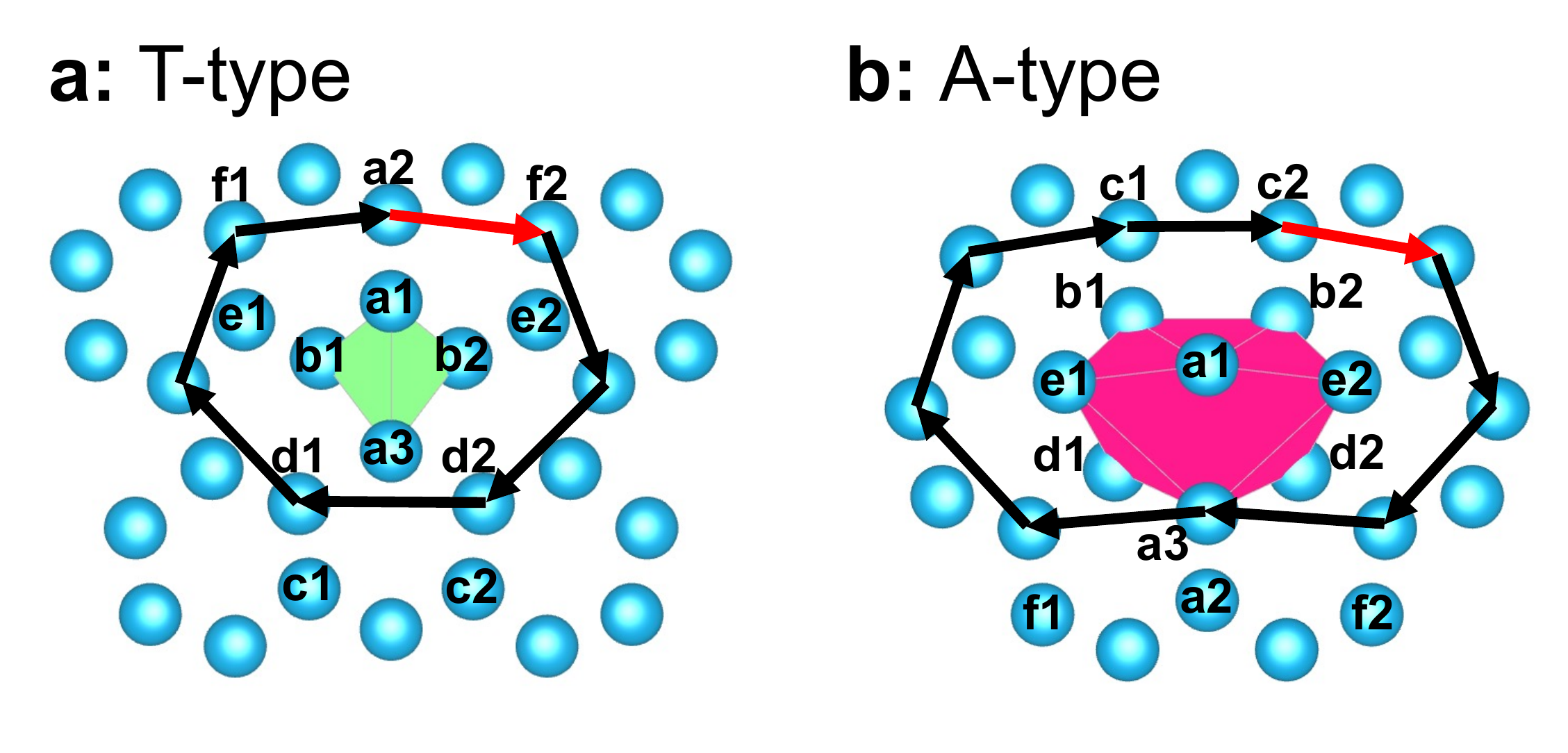}
\caption{Burgers circuit analysis for the (a) T-type and (b) A-type structural units. The black arrows show pairs of $\frac{1}{3}\langle2\overline{1}\overline{1}0\rangle$ vectors that are closed by the Burgers vectors $\vec{b}=\frac{1}{3}[2\overline{1}\overline{1}0]$ (red arrows). The nomenclature for the atomic columns is shown on top of them.}\label{loop}
\end{figure}

Three descriptors are employed for the decision making: 
1. The standard deviation of the horizontal coordinates of the three atomic sites \textbf{a1}, \textbf{a2}, and \textbf{a3} (see Fig.~\ref{loop} for the atomic site labels). A number close to zero corresponds to a straight GB plane; 
2. The absolute difference in $y$ coordinates between the \textbf{b1} and \textbf{b2} sites; 
3. The absolute difference in $y$ coordinates between the \textbf{d1} and \textbf{d2} sites. 
For the latter two descriptors, a number close to zero corresponds to a better mirror symmetry with respect to the GB plane. 
For a perfect T-type structure, all three descriptors have zero values, while the neighboring A-type motifs return bigger values and can be excluded. 
Likewise, for a perfect A-type structure, all three descriptors are zero, while the neighboring T-type motifs return bigger values and can be excluded. 
For decision making on the experimental images, the structural units with two or all of three descriptors returning smaller values (closer to zero) are designated (``Decision making'' in Figure~\ref{TtoA_Pattern}).

The union of the atom sites of the chosen matches describes most atoms at the grain boundary. 
To visualize the entire atomic grid topology, positions of atomic columns in the neighboring grains are still needed. 
For this purpose, we applied the motif extraction approach described in \cite{Alhasan2023} to determine the unit cell motifs of the left and right grains in the STEM images. 
The hexagons from both grains are then constructed and merged with the four-sided polygons from the detected T-type units and the eight-sided polygons from the detected A-type units to describe the atomic grid topology.

\subsection{STEM multi-slice image simulation}
\begin{figure}[h]%
\centering
\includegraphics[width=0.98\textwidth]{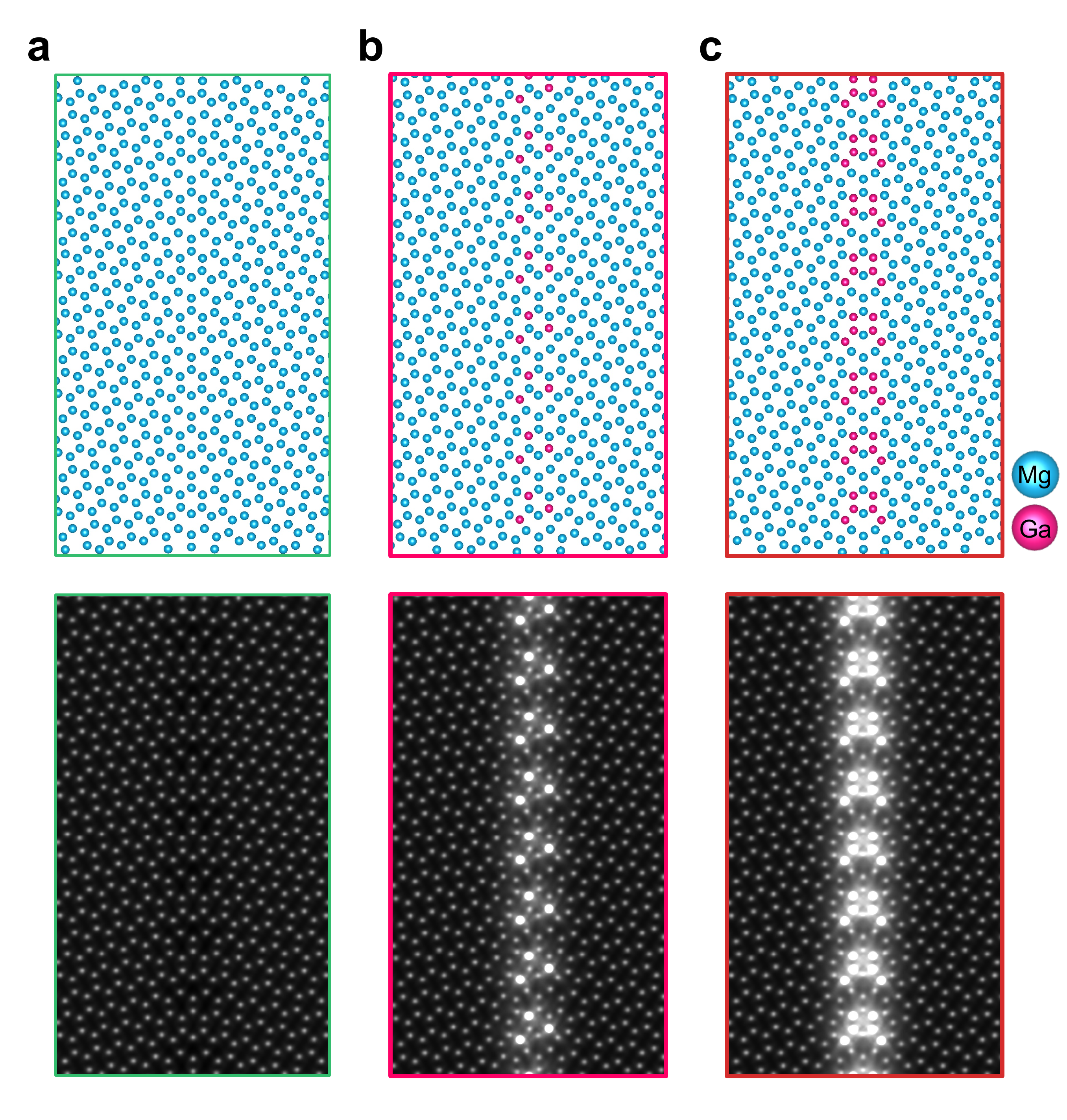}
\caption{Multislice STEM simulations for the structural models obtained from DFT calculations. Top: atomistic structural model. Bottom: Multislice STEM simulations. (a) T-type pure Mg $\Sigma7$ GB, and A-type units with (b) three and (c) six Ga columns.}\label{STEMsim}
\end{figure}

The purpose of the image simulation is to create simulated HAADF images based on the DFT calculated structures and compare them with the experimental measurements. 
We performed the STEM multi-slice simulations using the muSTEM (v5.2) software package \cite{Allen2015}. 
The microscope parameters for the simulations, such as the half-convergence angle (23.6~mard), primary electron energy (300~kV), and HAADF detector (62-200~mrad), were chosen according to the experimental conditions. 
Figure~\ref{STEMsim} displays the multislice STEM simulations of the structural models derived from DFT calculations. 
These models represent the T-type structural unit of pure Mg $\Sigma7$ GB, as well as the A-type units with 3-atom and 6-atom configurations.

\clearpage

\section{Extended Data: Figures}\label{SI}

\begin{figure}[h]%
\centering
\includegraphics[width=0.98\textwidth]{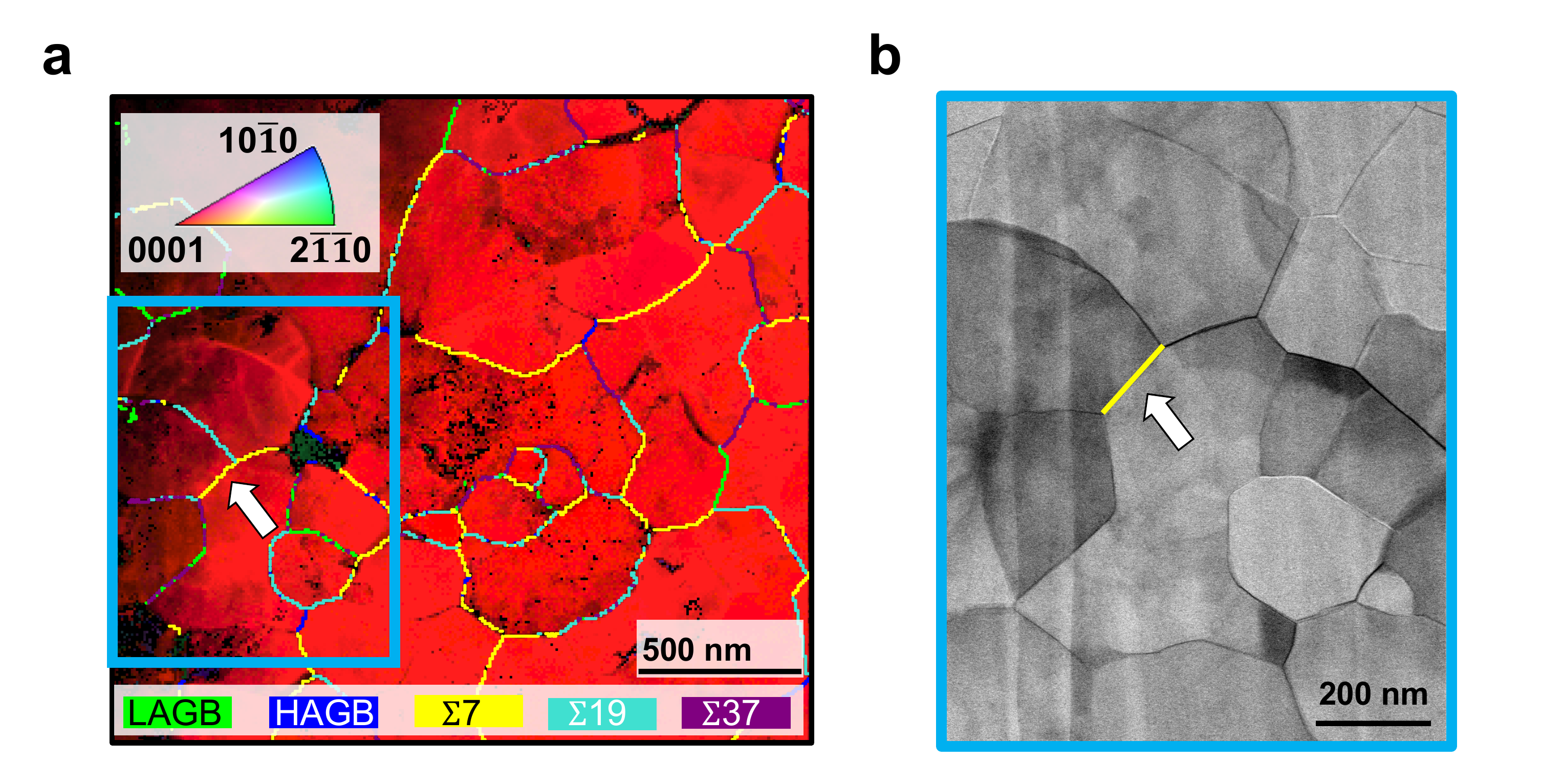}
\caption{(a) Orientation and GB maps reconstructed from the 4D-STEM data set. The thin film sample shows sharp (0001) texture (red color). Grains with a confidence index of less than 0.1 are shown in black. (b) Bright-field STEM image for the highlighted region in (a). White arrows in both figures point to the $\Sigma$7 GB for the high-resolution STEM study.}\label{4dSTEM}
\end{figure}

\begin{figure}[h]%
\centering
\includegraphics[width=0.8\textwidth]{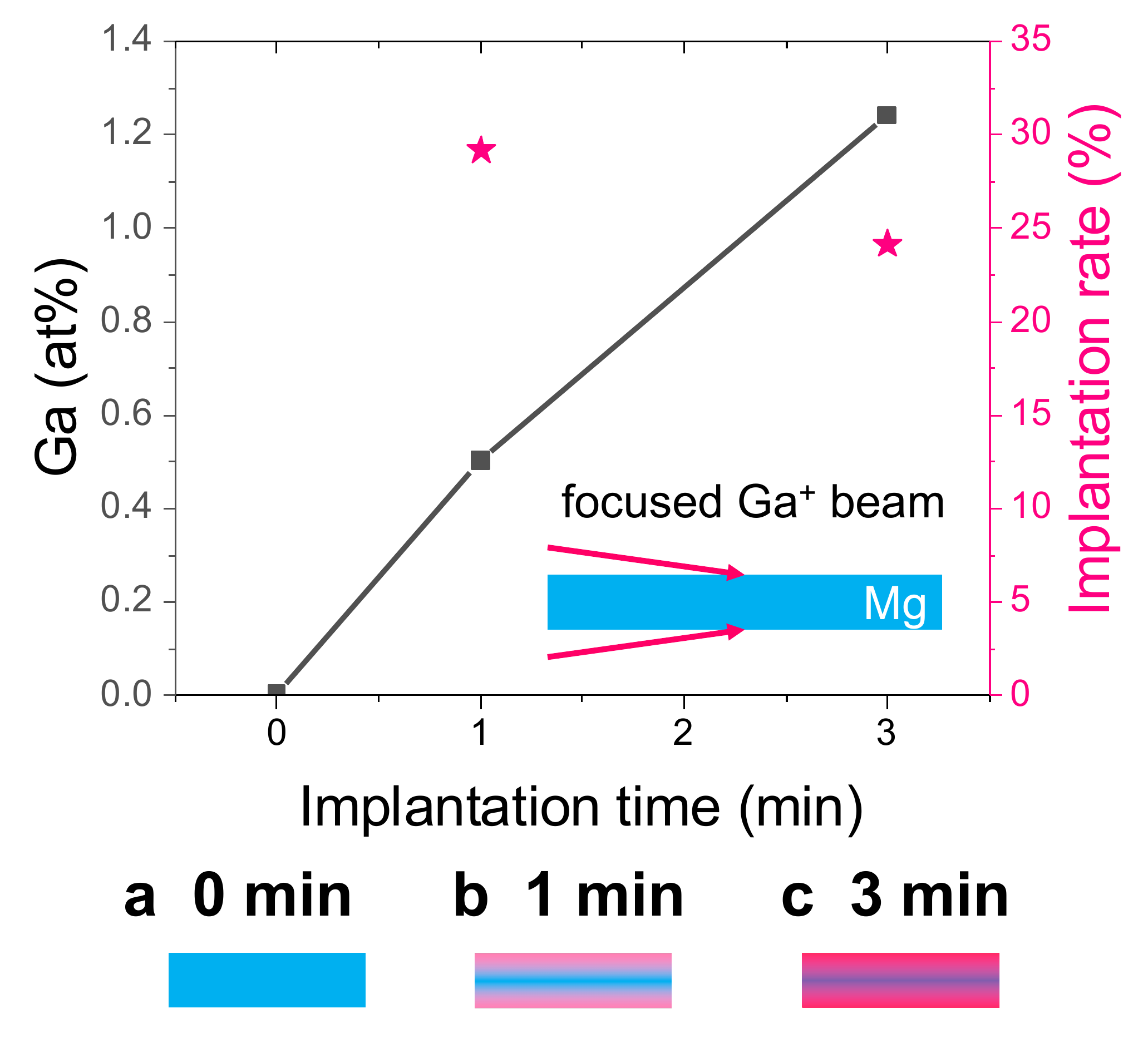}
\caption {The Ga composition inside the Mg sample as a function of the implantation time and the evaluated implantation rate. Schematics of Ga\textsuperscript{+} implantation by FIB is displayed in the inset at 5~kV, 7.7~pA, and grazing incidence of $\pm$8°. The three sample states of (a) pure Mg, after (b) 1~min and (c) 3~min implantation are schematically shown.}\label{implantation}
\end{figure}

\begin{figure}[h]%
\centering
\includegraphics[width=0.98\textwidth]{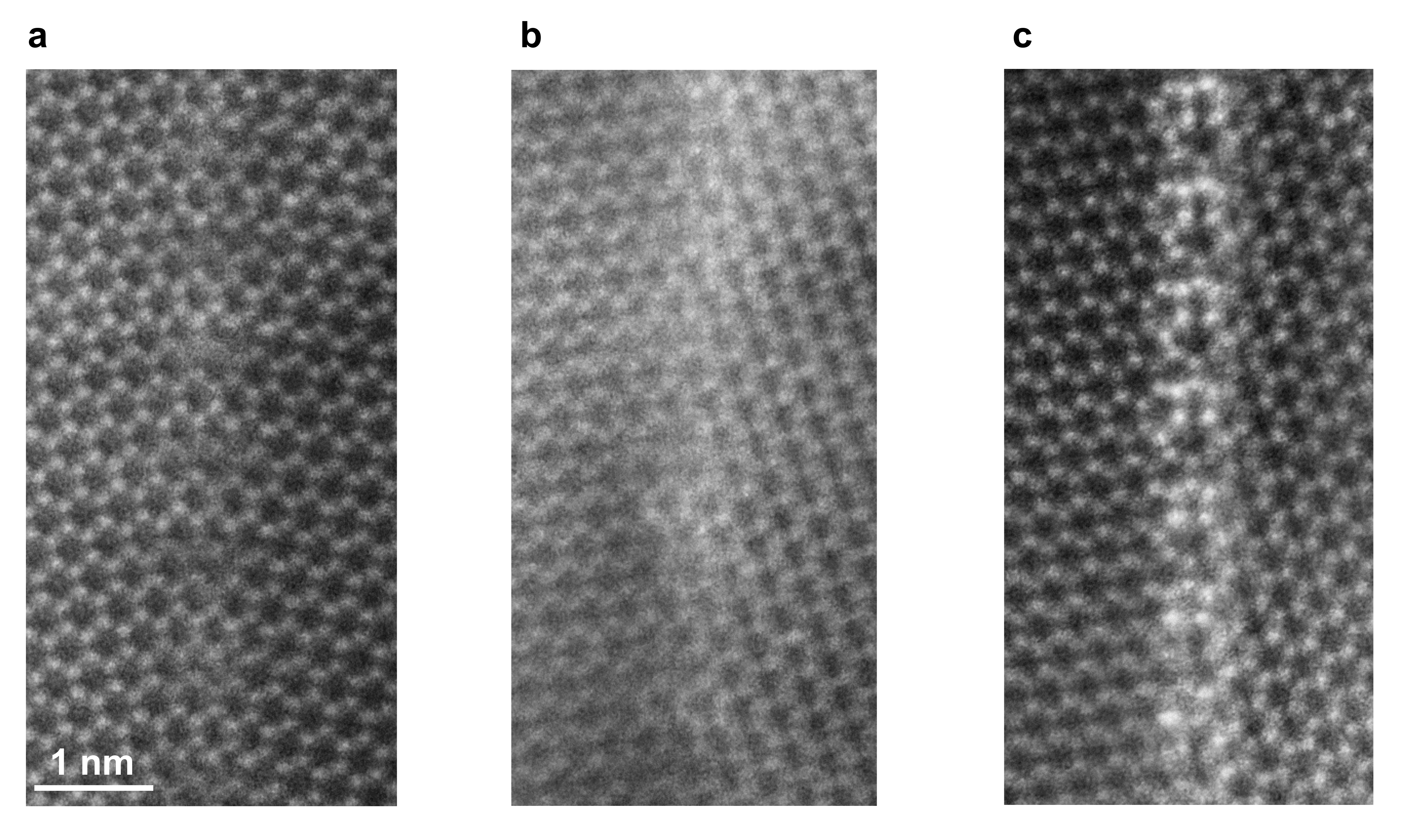}
\caption {HAADF-STEM images without overlaid grids, corresponding to the ones presented in Fig.~\ref{transformStructure}.}\label{transformStructure_SI}
\end{figure}

\begin{figure}[h]%
\centering
\includegraphics[width=0.98\textwidth]{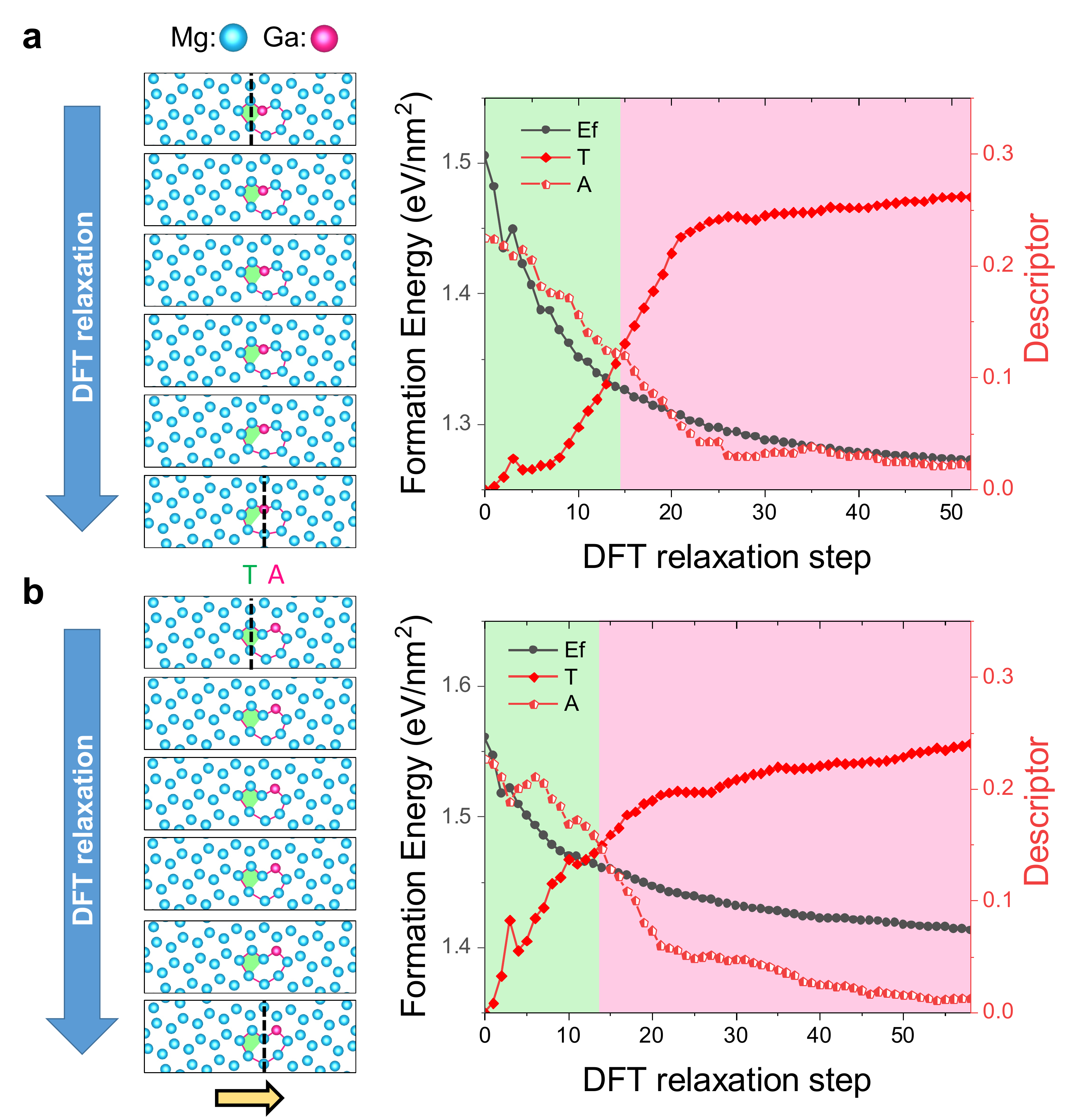}
\caption{Snapshots of DFT structural relaxation starting with a T-type unit with Ga atoms on the (a) \textbf{b2} and (b) \textbf{e2} sites, ending to an A-type unit with Ga atoms on the (a) \textbf{a1} and (b) \textbf{b2} sites. The left column of each image emphasizes the atomic structure at GBs, with the structural units of T-type and A-type distinguished by solid green shading and pink outline, respectively. The shift in GB plane is indicated by the arrow.
On the right column, the images display the evolution of formation energy as DFT relaxation steps. The descriptors to match T-type and A-type structural units are plotted in the same diagram. As a perfect match corresponds to a descriptor value of 0, the T-type and A-type structures are classified according to the lower value of their descriptors and shaded in green and pink colors, respectively.}\label{TtoA_DFT}
\end{figure}

\begin{figure}[h]%
\centering
\includegraphics[width=0.98\textwidth]{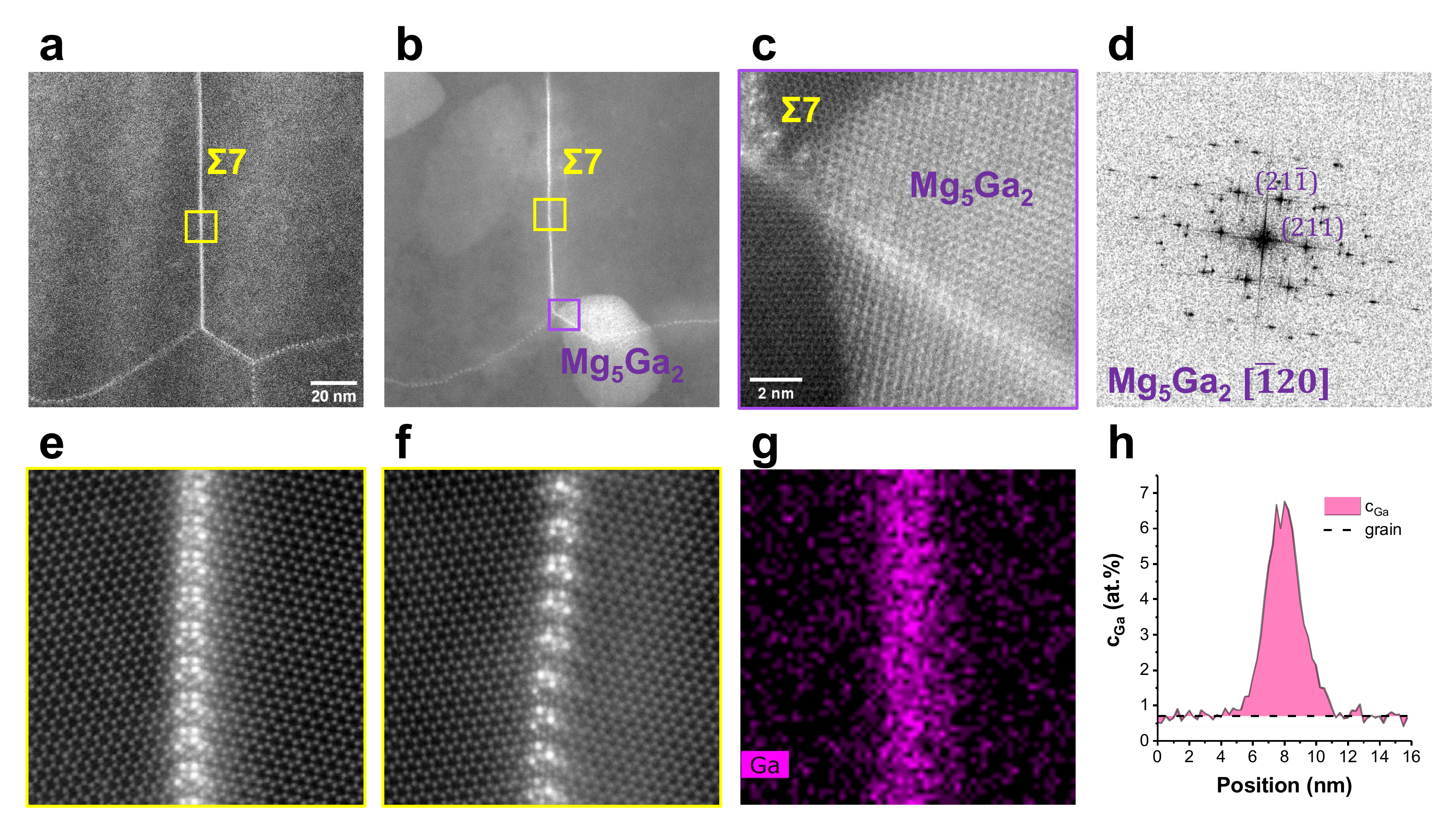}
\caption{HAADF-STEM images of the same $\Sigma$7 GB (a) 1 day and (b) 620 days after Ga\textsuperscript{+} beam thinning. The long storage time enabled longer range Ga diffusion to form bulk precipitates characterized as Mg\textsubscript{5}Ga\textsubscript{2} by (c) high resolution HAADF-STEM image and (d) the corresponding fast Fourier transformation. (e,f) High resolution HAADF-STEM images of the same $\Sigma$7 GB (e) 1 day and (f) 620 days after Ga\textsuperscript{+} beam thinning. (g) EDS Ga maps of the corresponding area in (f) and (h) Ga composition profile across the GB.}\label{transformTime_SI}
\end{figure}

\end{appendices}

\clearpage

\bibliography{bibliography.bib}{}


\begin{thebibliography}{70}
\ifx \bisbn   \undefined \def \bisbn  #1{ISBN #1}\fi
\ifx \binits  \undefined \def \binits#1{#1}\fi
\ifx \bauthor  \undefined \def \bauthor#1{#1}\fi
\ifx \batitle  \undefined \def \batitle#1{#1}\fi
\ifx \bjtitle  \undefined \def \bjtitle#1{#1}\fi
\ifx \bvolume  \undefined \def \bvolume#1{\textbf{#1}}\fi
\ifx \byear  \undefined \def \byear#1{#1}\fi
\ifx \bissue  \undefined \def \bissue#1{#1}\fi
\ifx \bfpage  \undefined \def \bfpage#1{#1}\fi
\ifx \blpage  \undefined \def \blpage #1{#1}\fi
\ifx \burl  \undefined \def \burl#1{\textsf{#1}}\fi
\ifx \doiurl  \undefined \def \doiurl#1{\url{https://doi.org/#1}}\fi
\ifx \betal  \undefined \def \betal{\textit{et al.}}\fi
\ifx \binstitute  \undefined \def \binstitute#1{#1}\fi
\ifx \binstitutionaled  \undefined \def \binstitutionaled#1{#1}\fi
\ifx \bctitle  \undefined \def \bctitle#1{#1}\fi
\ifx \beditor  \undefined \def \beditor#1{#1}\fi
\ifx \bpublisher  \undefined \def \bpublisher#1{#1}\fi
\ifx \bbtitle  \undefined \def \bbtitle#1{#1}\fi
\ifx \bedition  \undefined \def \bedition#1{#1}\fi
\ifx \bseriesno  \undefined \def \bseriesno#1{#1}\fi
\ifx \blocation  \undefined \def \blocation#1{#1}\fi
\ifx \bsertitle  \undefined \def \bsertitle#1{#1}\fi
\ifx \bsnm \undefined \def \bsnm#1{#1}\fi
\ifx \bsuffix \undefined \def \bsuffix#1{#1}\fi
\ifx \bparticle \undefined \def \bparticle#1{#1}\fi
\ifx \barticle \undefined \def \barticle#1{#1}\fi
\bibcommenthead
\ifx \bconfdate \undefined \def \bconfdate #1{#1}\fi
\ifx \botherref \undefined \def \botherref #1{#1}\fi
\ifx \url \undefined \def \url#1{\textsf{#1}}\fi
\ifx \bchapter \undefined \def \bchapter#1{#1}\fi
\ifx \bbook \undefined \def \bbook#1{#1}\fi
\ifx \bcomment \undefined \def \bcomment#1{#1}\fi
\ifx \oauthor \undefined \def \oauthor#1{#1}\fi
\ifx \citeauthoryear \undefined \def \citeauthoryear#1{#1}\fi
\ifx \endbibitem  \undefined \def \endbibitem {}\fi
\ifx \bconflocation  \undefined \def \bconflocation#1{#1}\fi
\ifx \arxivurl  \undefined \def \arxivurl#1{\textsf{#1}}\fi
\csname PreBibitemsHook\endcsname

\bibitem{Tasan2015}
\begin{barticle}
\bauthor{\bsnm{Tasan}, \binits{C.C.}},
\bauthor{\bsnm{Diehl}, \binits{M.}},
\bauthor{\bsnm{Yan}, \binits{D.}},
\bauthor{\bsnm{Bechtold}, \binits{M.}},
\bauthor{\bsnm{Roters}, \binits{F.}},
\bauthor{\bsnm{Schemmann}, \binits{L.}},
\bauthor{\bsnm{Zheng}, \binits{C.}},
\bauthor{\bsnm{Peranio}, \binits{N.}},
\bauthor{\bsnm{Ponge}, \binits{D.}},
\bauthor{\bsnm{Koyama}, \binits{M.}},
\bauthor{\bsnm{Tsuzaki}, \binits{K.}},
\bauthor{\bsnm{Raabe}, \binits{D.}}:
\batitle{{An Overview of Dual-Phase Steels: Advances in Microstructure-Oriented
  Processing and Micromechanically Guided Design}}.
\bjtitle{Annual Review of Materials Research}
\bvolume{45},
\bfpage{391}--\blpage{431}
(\byear{2015}).
\doiurl{10.1146/annurev-matsci-070214-021103}
\end{barticle}
\endbibitem

\bibitem{Sharma2009}
\begin{barticle}
\bauthor{\bsnm{Sharma}, \binits{A.}},
\bauthor{\bsnm{Tyagi}, \binits{V.V.}},
\bauthor{\bsnm{Chen}, \binits{C.R.}},
\bauthor{\bsnm{Buddhi}, \binits{D.}}:
\batitle{{Review on thermal energy storage with phase change materials and
  applications}}.
\bjtitle{Renewable and Sustainable Energy Reviews}
\bvolume{13}(\bissue{2}),
\bfpage{318}--\blpage{345}
(\byear{2009}).
\doiurl{10.1016/j.rser.2007.10.005}
\end{barticle}
\endbibitem

\bibitem{Raabe2007}
\begin{barticle}
\bauthor{\bsnm{Raabe}, \binits{D.}},
\bauthor{\bsnm{Sander}, \binits{B.}},
\bauthor{\bsnm{Fri{\'{a}}k}, \binits{M.}},
\bauthor{\bsnm{Ma}, \binits{D.}},
\bauthor{\bsnm{Neugebauer}, \binits{J.}}:
\batitle{{Theory-guided bottom-up design of $\beta$-titanium alloys as
  biomaterials based on first principles calculations: Theory and
  experiments}}.
\bjtitle{Acta Materialia}
\bvolume{55}(\bissue{13}),
\bfpage{4475}--\blpage{4487}
(\byear{2007}).
\doiurl{10.1016/j.actamat.2007.04.024}
\end{barticle}
\endbibitem

\bibitem{Gibbs1948}
\begin{botherref}
\oauthor{\bsnm{Gibbs}, \binits{J.W.}}:
The collected works of j. willard gibbs.
Technical report,
Yale Univ. Press,
(1948)
\end{botherref}
\endbibitem

\bibitem{Buban2006}
\begin{barticle}
\bauthor{\bsnm{Buban}, \binits{J.}},
\bauthor{\bsnm{Matsunaga}, \binits{K.}},
\bauthor{\bsnm{Chen}, \binits{J.}},
\bauthor{\bsnm{Shibata}, \binits{N.}},
\bauthor{\bsnm{Ching}, \binits{W.}},
\bauthor{\bsnm{Yamamoto}, \binits{T.}},
\bauthor{\bsnm{Ikuhara}, \binits{Y.}}:
\batitle{Grain boundary strengthening in alumina by rare earth impurities}.
\bjtitle{Science}
\bvolume{311}(\bissue{5758}),
\bfpage{212}--\blpage{215}
(\byear{2006})
\end{barticle}
\endbibitem

\bibitem{Lu2000}
\begin{barticle}
\bauthor{\bsnm{Lu}, \binits{L.}},
\bauthor{\bsnm{Sui}, \binits{M.}},
\bauthor{\bsnm{Lu}, \binits{K.}}:
\batitle{Superplastic extensibility of nanocrystalline copper at room
  temperature}.
\bjtitle{Science}
\bvolume{287}(\bissue{5457}),
\bfpage{1463}--\blpage{1466}
(\byear{2000})
\end{barticle}
\endbibitem

\bibitem{Legros2008}
\begin{barticle}
\bauthor{\bsnm{Legros}, \binits{M.}},
\bauthor{\bsnm{Dehm}, \binits{G.}},
\bauthor{\bsnm{Arzt}, \binits{E.}},
\bauthor{\bsnm{Balk}, \binits{T.J.}}:
\batitle{Observation of giant diffusivity along dislocation cores}.
\bjtitle{Science}
\bvolume{319}(\bissue{5870}),
\bfpage{1646}--\blpage{1649}
(\byear{2008})
\end{barticle}
\endbibitem

\bibitem{Lu2004}
\begin{barticle}
\bauthor{\bsnm{Lu}, \binits{L.}},
\bauthor{\bsnm{Shen}, \binits{Y.}},
\bauthor{\bsnm{Chen}, \binits{X.}},
\bauthor{\bsnm{Qian}, \binits{L.}},
\bauthor{\bsnm{Lu}, \binits{K.}}:
\batitle{Ultrahigh strength and high electrical conductivity in copper}.
\bjtitle{Science}
\bvolume{304}(\bissue{5669}),
\bfpage{422}--\blpage{426}
(\byear{2004})
\end{barticle}
\endbibitem

\bibitem{Cottrell1949}
\begin{barticle}
\bauthor{\bsnm{Cottrell}, \binits{A.H.}},
\bauthor{\bsnm{Bilby}, \binits{B.A.}}:
\batitle{{Dislocation theory of yielding and strain ageing of iron}}.
\bjtitle{Proceedings of the Physical Society. Section A}
\bvolume{62}(\bissue{1}),
\bfpage{49}--\blpage{62}
(\byear{1949}).
\doiurl{10.1088/0370-1298/62/1/308}
\end{barticle}
\endbibitem

\bibitem{Kuzmina2015}
\begin{barticle}
\bauthor{\bsnm{Kuzmina}, \binits{M.}},
\bauthor{\bsnm{Herbig}, \binits{M.}},
\bauthor{\bsnm{Ponge}, \binits{D.}},
\bauthor{\bsnm{Sandl{\"{o}}bes}, \binits{S.}},
\bauthor{\bsnm{Raabe}, \binits{D.}}:
\batitle{{Linear complexions: Confined chemical and structural states at
  dislocations}}.
\bjtitle{Science}
\bvolume{349}(\bissue{6252}),
\bfpage{1080}--\blpage{1083}
(\byear{2015}).
\doiurl{10.1126/science.aab2633}
\end{barticle}
\endbibitem

\bibitem{Zhou2021}
\begin{barticle}
\bauthor{\bsnm{Zhou}, \binits{X.}},
\bauthor{\bsnm{Mianroodi}, \binits{J.R.}},
\bauthor{\bparticle{Kwiatkowski~da} \bsnm{Silva}, \binits{A.}},
\bauthor{\bsnm{Koenig}, \binits{T.}},
\bauthor{\bsnm{Thompson}, \binits{G.B.}},
\bauthor{\bsnm{Shanthraj}, \binits{P.}},
\bauthor{\bsnm{Ponge}, \binits{D.}},
\bauthor{\bsnm{Gault}, \binits{B.}},
\bauthor{\bsnm{Svendsen}, \binits{B.}},
\bauthor{\bsnm{Raabe}, \binits{D.}}:
\batitle{The hidden structure dependence of the chemical life of dislocations}.
\bjtitle{Science Advances}
\bvolume{7}(\bissue{16}),
\bfpage{0563}
(\byear{2021})
\end{barticle}
\endbibitem

\bibitem{Yu2022}
\begin{botherref}
\oauthor{\bsnm{Yu}, \binits{Y.}},
\oauthor{\bsnm{Zhou}, \binits{C.}},
\oauthor{\bsnm{Zhang}, \binits{X.}},
\oauthor{\bsnm{Abdellaoui}, \binits{L.}},
\oauthor{\bsnm{Doberstein}, \binits{C.}},
\oauthor{\bsnm{Berkels}, \binits{B.}},
\oauthor{\bsnm{Ge}, \binits{B.}},
\oauthor{\bsnm{Qiao}, \binits{G.}},
\oauthor{\bsnm{Scheu}, \binits{C.}},
\oauthor{\bsnm{Wuttig}, \binits{M.}},
\oauthor{\bsnm{Cojocaru-Mir{\'{e}}din}, \binits{O.}},
\oauthor{\bsnm{Zhang}, \binits{S.}}:
{Dynamic doping and Cottrell atmosphere optimize the thermoelectric performance
  of n-type PbTe over a broad temperature interval}.
Nano Energy
\textbf{101}(April)
(2022).
\doiurl{10.1016/j.nanoen.2022.107576}
\end{botherref}
\endbibitem

\bibitem{Suzuki1962}
\begin{barticle}
\bauthor{\bsnm{Suzuki}, \binits{H.}}:
\batitle{Segregation of solute atoms to stacking faults}.
\bjtitle{Journal of the Physical Society of Japan}
\bvolume{17}(\bissue{2}),
\bfpage{322}--\blpage{325}
(\byear{1962}).
\doiurl{10.1143/JPSJ.17.322}
\end{barticle}
\endbibitem

\bibitem{Palanisamy2019}
\begin{barticle}
\bauthor{\bsnm{Palanisamy}, \binits{D.}},
\bauthor{\bsnm{Raabe}, \binits{D.}},
\bauthor{\bsnm{Gault}, \binits{B.}}:
\batitle{On the compositional partitioning during phase transformation in a
  binary ferromagnetic mnal alloy}.
\bjtitle{Acta Materialia}
\bvolume{174},
\bfpage{227}--\blpage{236}
(\byear{2019})
\end{barticle}
\endbibitem

\bibitem{Lejcek1991}
\begin{barticle}
\bauthor{\bsnm{Lej{\v{c}}ek}, \binits{P.}},
\bauthor{\bsnm{Hofmann}, \binits{S.}}:
\batitle{Segregation enthalpies of phosphorus, carbon and silicon at
  $\{$013$\}$ and $\{$012$\}$ symmetrical tilt grain boundaries in an fe-3.5
  at.\% si alloy}.
\bjtitle{Acta metallurgica et materialia}
\bvolume{39}(\bissue{10}),
\bfpage{2469}--\blpage{2476}
(\byear{1991})
\end{barticle}
\endbibitem

\bibitem{Kirchheim2002}
\begin{barticle}
\bauthor{\bsnm{Kirchheim}, \binits{R.}}:
\batitle{Grain coarsening inhibited by solute segregation}.
\bjtitle{Acta Materialia}
\bvolume{50}(\bissue{2}),
\bfpage{413}--\blpage{419}
(\byear{2002})
\end{barticle}
\endbibitem

\bibitem{Wang2011}
\begin{barticle}
\bauthor{\bsnm{Wang}, \binits{Z.}},
\bauthor{\bsnm{Saito}, \binits{M.}},
\bauthor{\bsnm{McKenna}, \binits{K.P.}},
\bauthor{\bsnm{Gu}, \binits{L.}},
\bauthor{\bsnm{Tsukimoto}, \binits{S.}},
\bauthor{\bsnm{Shluger}, \binits{A.L.}},
\bauthor{\bsnm{Ikuhara}, \binits{Y.}}:
\batitle{Atom-resolved imaging of ordered defect superstructures at individual
  grain boundaries}.
\bjtitle{Nature}
\bvolume{479}(\bissue{7373}),
\bfpage{380}--\blpage{383}
(\byear{2011})
\end{barticle}
\endbibitem

\bibitem{Schuh2012}
\begin{barticle}
\bauthor{\bsnm{Chookajorn}, \binits{T.}},
\bauthor{\bsnm{Murdoch}, \binits{H.A.}},
\bauthor{\bsnm{Schuh}, \binits{C.A.}}:
\batitle{Design of stable nanocrystalline alloys}.
\bjtitle{Science}
\bvolume{337}(\bissue{6097}),
\bfpage{951}--\blpage{954}
(\byear{2012}).
\doiurl{10.1126/science.1224737}
\end{barticle}
\endbibitem

\bibitem{Nie2013}
\begin{barticle}
\bauthor{\bsnm{Nie}, \binits{J.F.}},
\bauthor{\bsnm{Zhu}, \binits{Y.}},
\bauthor{\bsnm{Liu}, \binits{J.}},
\bauthor{\bsnm{Fang}, \binits{X.-Y.}}:
\batitle{Periodic segregation of solute atoms in fully coherent twin
  boundaries}.
\bjtitle{Science}
\bvolume{340}(\bissue{6135}),
\bfpage{957}--\blpage{960}
(\byear{2013})
\end{barticle}
\endbibitem

\bibitem{Raabe2014}
\begin{barticle}
\bauthor{\bsnm{Raabe}, \binits{D.}},
\bauthor{\bsnm{Herbig}, \binits{M.}},
\bauthor{\bsnm{Sandl{\"{o}}bes}, \binits{S.}},
\bauthor{\bsnm{Li}, \binits{Y.}},
\bauthor{\bsnm{Tytko}, \binits{D.}},
\bauthor{\bsnm{Kuzmina}, \binits{M.}},
\bauthor{\bsnm{Ponge}, \binits{D.}},
\bauthor{\bsnm{Choi}, \binits{P.P.}}:
\batitle{{Grain boundary segregation engineering in metallic alloys: A pathway
  to the design of interfaces}}.
\bjtitle{Current Opinion in Solid State and Materials Science}
\bvolume{18}(\bissue{4}),
\bfpage{253}--\blpage{261}
(\byear{2014}).
\doiurl{10.1016/j.cossms.2014.06.002}
\end{barticle}
\endbibitem

\bibitem{Yu2017}
\begin{barticle}
\bauthor{\bsnm{Yu}, \binits{Z.}},
\bauthor{\bsnm{Cantwell}, \binits{P.R.}},
\bauthor{\bsnm{Gao}, \binits{Q.}},
\bauthor{\bsnm{Yin}, \binits{D.}},
\bauthor{\bsnm{Zhang}, \binits{Y.}},
\bauthor{\bsnm{Zhou}, \binits{N.}},
\bauthor{\bsnm{Rohrer}, \binits{G.S.}},
\bauthor{\bsnm{Widom}, \binits{M.}},
\bauthor{\bsnm{Luo}, \binits{J.}},
\bauthor{\bsnm{Harmer}, \binits{M.P.}}:
\batitle{Segregation-induced ordered superstructures at general grain
  boundaries in a nickel-bismuth alloy}.
\bjtitle{Science}
\bvolume{358}(\bissue{6359}),
\bfpage{97}--\blpage{101}
(\byear{2017})
\end{barticle}
\endbibitem

\bibitem{Lejcek2017}
\begin{barticle}
\bauthor{\bsnm{Lej{\v{c}}ek}, \binits{P.}},
\bauthor{\bsnm{{\v{S}}ob}, \binits{M.}},
\bauthor{\bsnm{Paidar}, \binits{V.}}:
\batitle{Interfacial segregation and grain boundary embrittlement: An overview
  and critical assessment of experimental data and calculated results}.
\bjtitle{Progress in Materials Science}
\bvolume{87},
\bfpage{83}--\blpage{139}
(\byear{2017})
\end{barticle}
\endbibitem

\bibitem{Zhou2023}
\begin{barticle}
\bauthor{\bsnm{Zhou}, \binits{X.}},
\bauthor{\bsnm{Ahmadian}, \binits{A.}},
\bauthor{\bsnm{Gault}, \binits{B.}},
\bauthor{\bsnm{Ophus}, \binits{C.}},
\bauthor{\bsnm{Liebscher}, \binits{C.H.}},
\bauthor{\bsnm{Dehm}, \binits{G.}},
\bauthor{\bsnm{Raabe}, \binits{D.}}:
\batitle{Atomic motifs govern the decoration of grain boundaries by
  interstitial solutes}.
\bjtitle{Nature Communications}
\bvolume{14}(\bissue{1}),
\bfpage{3535}
(\byear{2023})
\end{barticle}
\endbibitem

\bibitem{Hart1968}
\begin{barticle}
\bauthor{\bsnm{Hart}, \binits{E.W.}}:
\batitle{Two-dimensional phase transformation in grain boundaries}.
\bjtitle{Scripta Metallurgica}
\bvolume{2}(\bissue{3}),
\bfpage{179}--\blpage{182}
(\byear{1968}).
\doiurl{10.1016/0036-9748(68)90222-6}
\end{barticle}
\endbibitem

\bibitem{Frolov2015}
\begin{barticle}
\bauthor{\bsnm{Frolov}, \binits{T.}},
\bauthor{\bsnm{Mishin}, \binits{Y.}}:
\batitle{Phases, phase equilibria, and phase rules in low-dimensional systems}.
\bjtitle{The Journal of chemical physics}
\bvolume{143}(\bissue{4}),
\bfpage{044706}
(\byear{2015})
\end{barticle}
\endbibitem

\bibitem{Brink2022}
\begin{botherref}
\oauthor{\bsnm{Brink}, \binits{T.}},
\oauthor{\bsnm{Langenohl}, \binits{L.}},
\oauthor{\bsnm{Bishara}, \binits{H.}},
\oauthor{\bsnm{Dehm}, \binits{G.}}:
{Universality of grain boundary phases in fcc metals: Case study on high-angle
  [111] symmetric tilt grain boundaries}.
Physical Review B - Condensed Matter and Materials Physics
\textbf{054103}
(2022)
{\href{https://arxiv.org/abs/2211.14170}{{arXiv:2211.14170}}}.
\doiurl{10.1103/PhysRevB.107.054103}
\end{botherref}
\endbibitem

\bibitem{Dillon2007}
\begin{barticle}
\bauthor{\bsnm{Dillon}, \binits{S.J.}},
\bauthor{\bsnm{Tang}, \binits{M.}},
\bauthor{\bsnm{Carter}, \binits{W.C.}},
\bauthor{\bsnm{Harmer}, \binits{M.P.}}:
\batitle{{Complexion: A new concept for kinetic engineering in materials
  science}}.
\bjtitle{Acta Materialia}
\bvolume{55}(\bissue{18}),
\bfpage{6208}--\blpage{6218}
(\byear{2007}).
\doiurl{10.1016/j.actamat.2007.07.029}
\end{barticle}
\endbibitem

\bibitem{Harmer2011}
\begin{barticle}
\bauthor{\bsnm{Harmer}, \binits{M.P.}}:
\batitle{The phase behavior of interfaces}.
\bjtitle{Science}
\bvolume{332}(\bissue{6026}),
\bfpage{182}--\blpage{183}
(\byear{2011})
\end{barticle}
\endbibitem

\bibitem{Luo2011}
\begin{barticle}
\bauthor{\bsnm{Luo}, \binits{J.}},
\bauthor{\bsnm{Cheng}, \binits{H.}},
\bauthor{\bsnm{Asl}, \binits{K.M.}},
\bauthor{\bsnm{Kiely}, \binits{C.J.}},
\bauthor{\bsnm{Harmer}, \binits{M.P.}}:
\batitle{The role of a bilayer interfacial phase on liquid metal
  embrittlement}.
\bjtitle{Science}
\bvolume{333}(\bissue{6050}),
\bfpage{1730}--\blpage{1733}
(\byear{2011})
\end{barticle}
\endbibitem

\bibitem{Kaplan2013}
\begin{barticle}
\bauthor{\bsnm{Kaplan}, \binits{W.D.}},
\bauthor{\bsnm{Chatain}, \binits{D.}},
\bauthor{\bsnm{Wynblatt}, \binits{P.}},
\bauthor{\bsnm{Carter}, \binits{W.C.}}:
\batitle{{A review of wetting versus adsorption, complexions, and related
  phenomena: The rosetta stone of wetting}}.
\bjtitle{Journal of Materials Science}
\bvolume{48}(\bissue{17}),
\bfpage{5681}--\blpage{5717}
(\byear{2013}).
\doiurl{10.1007/s10853-013-7462-y}
\end{barticle}
\endbibitem

\bibitem{Cantwell2014}
\begin{barticle}
\bauthor{\bsnm{Cantwell}, \binits{P.R.}},
\bauthor{\bsnm{Tang}, \binits{M.}},
\bauthor{\bsnm{Dillon}, \binits{S.J.}},
\bauthor{\bsnm{Luo}, \binits{J.}},
\bauthor{\bsnm{Rohrer}, \binits{G.S.}},
\bauthor{\bsnm{Harmer}, \binits{M.P.}}:
\batitle{{Grain boundary complexions}}.
\bjtitle{Acta Materialia}
\bvolume{62}(\bissue{1}),
\bfpage{1}--\blpage{48}
(\byear{2014}).
\doiurl{10.1016/j.actamat.2013.07.037}
\end{barticle}
\endbibitem

\bibitem{Korte-Kerzel2022}
\begin{barticle}
\bauthor{\bsnm{Korte-Kerzel}, \binits{S.}},
\bauthor{\bsnm{Hickel}, \binits{T.}},
\bauthor{\bsnm{Huber}, \binits{L.}},
\bauthor{\bsnm{Raabe}, \binits{D.}},
\bauthor{\bsnm{Sandl{\"{o}}bes-Haut}, \binits{S.}},
\bauthor{\bsnm{Todorova}, \binits{M.}},
\bauthor{\bsnm{Neugebauer}, \binits{J.}}:
\batitle{{Defect phases–thermodynamics and impact on material properties}}.
\bjtitle{International Materials Reviews}
\bvolume{67}(\bissue{1}),
\bfpage{89}--\blpage{117}
(\byear{2022}).
\doiurl{10.1080/09506608.2021.1930734}
\end{barticle}
\endbibitem

\bibitem{Bishara2021}
\begin{barticle}
\bauthor{\bsnm{Bishara}, \binits{H.}},
\bauthor{\bsnm{Lee}, \binits{S.}},
\bauthor{\bsnm{Brink}, \binits{T.}},
\bauthor{\bsnm{Ghidelli}, \binits{M.}},
\bauthor{\bsnm{Dehm}, \binits{G.}}:
\batitle{{Understanding Grain Boundary Electrical Resistivity in Cu: The Effect
  of Boundary Structure}}.
\bjtitle{ACS Nano}
\bvolume{15}(\bissue{10}),
\bfpage{16607}--\blpage{16615}
(\byear{2021}).
\doiurl{10.1021/acsnano.1c06367}
\end{barticle}
\endbibitem

\bibitem{BuenoVilloro2023}
\begin{barticle}
\bauthor{\bsnm{{Bueno Villoro}}, \binits{R.}},
\bauthor{\bsnm{Zavanelli}, \binits{D.}},
\bauthor{\bsnm{Jung}, \binits{C.}},
\bauthor{\bsnm{Mattlat}, \binits{D.A.}},
\bauthor{\bsnm{{Hatami Naderloo}}, \binits{R.}},
\bauthor{\bsnm{P{\'{e}}rez}, \binits{N.}},
\bauthor{\bsnm{Nielsch}, \binits{K.}},
\bauthor{\bsnm{Snyder}, \binits{G.J.}},
\bauthor{\bsnm{Scheu}, \binits{C.}},
\bauthor{\bsnm{He}, \binits{R.}},
\bauthor{\bsnm{Zhang}, \binits{S.}}:
\batitle{{Grain Boundary Phases in NbFeSb Half-Heusler Alloys: A New Avenue to
  Tune Transport Properties of Thermoelectric Materials}}.
\bjtitle{Advanced Energy Materials}
(\byear{2023}).
\doiurl{10.1002/aenm.202204321}
\end{barticle}
\endbibitem

\bibitem{BuenoVilloro2023a}
\begin{barticle}
\bauthor{\bsnm{{Bueno Villoro}}, \binits{R.}},
\bauthor{\bsnm{Wood}, \binits{M.}},
\bauthor{\bsnm{Luo}, \binits{T.}},
\bauthor{\bsnm{Bishara}, \binits{H.}},
\bauthor{\bsnm{Abdellaoui}, \binits{L.}},
\bauthor{\bsnm{Zavanelli}, \binits{D.}},
\bauthor{\bsnm{Gault}, \binits{B.}},
\bauthor{\bsnm{Snyder}, \binits{G.J.}},
\bauthor{\bsnm{Scheu}, \binits{C.}},
\bauthor{\bsnm{Zhang}, \binits{S.}}:
\batitle{{Fe Segregation as a Tool to Enhance Electrical Conductivity of Grain
  Boundaries in Ti ( Co , Fe ) Sb Half Heusler Thermoelectrics}}.
\bjtitle{Acta Materialia}
\bvolume{249}(\bissue{March}),
\bfpage{118816}
(\byear{2023}).
\doiurl{10.1016/j.actamat.2023.118816}
\end{barticle}
\endbibitem

\bibitem{Duerrschnabel2017}
\begin{barticle}
\bauthor{\bsnm{Duerrschnabel}, \binits{M.}},
\bauthor{\bsnm{Yi}, \binits{M.}},
\bauthor{\bsnm{Uestuener}, \binits{K.}},
\bauthor{\bsnm{Liesegang}, \binits{M.}},
\bauthor{\bsnm{Katter}, \binits{M.}},
\bauthor{\bsnm{Kleebe}, \binits{H.J.}},
\bauthor{\bsnm{Xu}, \binits{B.}},
\bauthor{\bsnm{Gutfleisch}, \binits{O.}},
\bauthor{\bsnm{Molina-Luna}, \binits{L.}}:
\batitle{{Atomic structure and domain wall pinning in samarium-cobalt-based
  permanent magnets}}.
\bjtitle{Nature Communications}
\bvolume{8}(\bissue{1}),
\bfpage{1}--\blpage{7}
(\byear{2017}).
\doiurl{10.1038/s41467-017-00059-9}
\end{barticle}
\endbibitem

\bibitem{Hall1951}
\begin{barticle}
\bauthor{\bsnm{Hall}, \binits{E.O.}}:
\batitle{{The deformation and ageing of mild steel: III Discussion of
  results}}.
\bjtitle{Proceedings of the Physical Society. Section B}
\bvolume{64}(\bissue{9}),
\bfpage{747}--\blpage{753}
(\byear{1951}).
\doiurl{10.1088/0370-1301/64/9/303}
\end{barticle}
\endbibitem

\bibitem{Wu1994}
\begin{barticle}
\bauthor{\bsnm{Wu}, \binits{R.}},
\bauthor{\bsnm{Freeman}, \binits{A.}},
\bauthor{\bsnm{Olson}, \binits{G.B.}}:
\batitle{First principles determination of the effects of phosphorus and boron
  on iron grain boundary cohesion}.
\bjtitle{Science}
\bvolume{265}(\bissue{5170}),
\bfpage{376}--\blpage{380}
(\byear{1994})
\end{barticle}
\endbibitem

\bibitem{Khalajhedayati2016}
\begin{barticle}
\bauthor{\bsnm{Khalajhedayati}, \binits{A.}},
\bauthor{\bsnm{Pan}, \binits{Z.}},
\bauthor{\bsnm{Rupert}, \binits{T.J.}}:
\batitle{Manipulating the interfacial structure of nanomaterials to achieve a
  unique combination of strength and ductility}.
\bjtitle{Nature communications}
\bvolume{7}(\bissue{1}),
\bfpage{10802}
(\byear{2016})
\end{barticle}
\endbibitem

\bibitem{Krause2018}
\begin{barticle}
\bauthor{\bsnm{Krause}, \binits{A.R.}},
\bauthor{\bsnm{Cantwell}, \binits{P.R.}},
\bauthor{\bsnm{Marvel}, \binits{C.J.}},
\bauthor{\bsnm{Compson}, \binits{C.}},
\bauthor{\bsnm{Rickman}, \binits{J.M.}},
\bauthor{\bsnm{Harmer}, \binits{M.P.}}:
\batitle{{Review of grain boundary complexion engineering: Know your
  boundaries}}.
\bjtitle{Journal of the American Ceramic Society}
\bvolume{102}(\bissue{2}),
\bfpage{778}--\blpage{800}
(\byear{2018}).
\doiurl{10.1111/jace.16045}
\end{barticle}
\endbibitem

\bibitem{Cantwell2020}
\begin{barticle}
\bauthor{\bsnm{Cantwell}, \binits{P.R.}},
\bauthor{\bsnm{Frolov}, \binits{T.}},
\bauthor{\bsnm{Rupert}, \binits{T.J.}},
\bauthor{\bsnm{Krause}, \binits{A.R.}},
\bauthor{\bsnm{Marvel}, \binits{C.J.}},
\bauthor{\bsnm{Rohrer}, \binits{G.S.}},
\bauthor{\bsnm{Rickman}, \binits{J.M.}},
\bauthor{\bsnm{Harmer}, \binits{M.P.}}:
\batitle{{Grain Boundary Complexion Transitions}}.
\bjtitle{Annual Review of Materials Research}
\bvolume{50},
\bfpage{465}--\blpage{492}
(\byear{2020}).
\doiurl{10.1146/annurev-matsci-081619-114055}
\end{barticle}
\endbibitem

\bibitem{Dehm2022}
\begin{barticle}
\bauthor{\bsnm{Dehm}, \binits{G.}},
\bauthor{\bsnm{Cairney}, \binits{J.}}:
\batitle{{Implication of grain-boundary structure and chemistry on plasticity
  and failure}}.
\bjtitle{MRS Bulletin}
\bvolume{47}(\bissue{8}),
\bfpage{800}--\blpage{807}
(\byear{2022}).
\doiurl{10.1557/s43577-022-00378-3}
\end{barticle}
\endbibitem

\bibitem{Cahn1982}
\begin{barticle}
\bauthor{\bsnm{{Cahn, J. W.}}}:
\batitle{Transitions and phase equilibria among grain boundary structures}.
\bjtitle{J. Phys. Colloques}
\bvolume{43}(\bissue{C6}),
\bfpage{6}--\blpage{1996213}
(\byear{1982}).
\doiurl{10.1051/jphyscol:1982619}
\end{barticle}
\endbibitem

\bibitem{Rottman1991}
\begin{barticle}
\bauthor{\bsnm{Rottman}, \binits{C.}}:
\batitle{Phase transitions at internal interfaces}.
\bjtitle{MRS Proceedings}
\bvolume{238},
\bfpage{191}
(\byear{1991}).
\doiurl{10.1557/PROC-238-191}
\end{barticle}
\endbibitem

\bibitem{Tang2006}
\begin{barticle}
\bauthor{\bsnm{Tang}, \binits{M.}},
\bauthor{\bsnm{Carter}, \binits{W.C.}},
\bauthor{\bsnm{Cannon}, \binits{R.M.}}:
\batitle{Diffuse interface model for structural transitions of grain
  boundaries}.
\bjtitle{Phys. Rev. B}
\bvolume{73},
\bfpage{024102}
(\byear{2006}).
\doiurl{10.1103/PhysRevB.73.024102}
\end{barticle}
\endbibitem

\bibitem{Frolov2012}
\begin{barticle}
\bauthor{\bsnm{Frolov}, \binits{T.}},
\bauthor{\bsnm{Mishin}, \binits{Y.}}:
\batitle{{Thermodynamics of coherent interfaces under mechanical stresses. I.
  Theory}}.
\bjtitle{Physical Review B - Condensed Matter and Materials Physics}
\bvolume{85}(\bissue{22}),
\bfpage{12}--\blpage{15}
(\byear{2012})
{\href{https://arxiv.org/abs/1304.0144}{{arXiv:1304.0144}}}.
\doiurl{10.1103/PhysRevB.85.224106}
\end{barticle}
\endbibitem

\bibitem{Frolov2012a}
\begin{botherref}
\oauthor{\bsnm{Frolov}, \binits{T.}},
\oauthor{\bsnm{Mishin}, \binits{Y.}}:
{Thermodynamics of coherent interfaces under mechanical stresses. II.
  Application to atomistic simulation of grain boundaries}.
Physical Review B - Condensed Matter and Materials Physics
\textbf{85}(22)
(2012).
\doiurl{10.1103/PhysRevB.85.224107}
\end{botherref}
\endbibitem

\bibitem{Frolov2013}
\begin{barticle}
\bauthor{\bsnm{Frolov}, \binits{T.}},
\bauthor{\bsnm{Divinski}, \binits{S.V.}},
\bauthor{\bsnm{Asta}, \binits{M.}},
\bauthor{\bsnm{Mishin}, \binits{Y.}}:
\batitle{{Effect of interface phase transformations on diffusion and
  segregation in high-angle grain boundaries}}.
\bjtitle{Physical Review Letters}
\bvolume{110}(\bissue{25}),
\bfpage{1}--\blpage{5}
(\byear{2013})
{\href{https://arxiv.org/abs/1304.0276}{{arXiv:1304.0276}}}.
\doiurl{10.1103/PhysRevLett.110.255502}
\end{barticle}
\endbibitem

\bibitem{Meiners2020}
\begin{barticle}
\bauthor{\bsnm{Meiners}, \binits{T.}},
\bauthor{\bsnm{Frolov}, \binits{T.}},
\bauthor{\bsnm{Rudd}, \binits{R.E.}},
\bauthor{\bsnm{Dehm}, \binits{G.}},
\bauthor{\bsnm{Liebscher}, \binits{C.H.}}:
\batitle{{Observations of grain-boundary phase transformations in an elemental
  metal}}.
\bjtitle{Nature}
\bvolume{579}(\bissue{7799}),
\bfpage{375}--\blpage{378}
(\byear{2020}).
\doiurl{10.1038/s41586-020-2082-6}
\end{barticle}
\endbibitem

\bibitem{Frommeyer2022}
\begin{barticle}
\bauthor{\bsnm{Frommeyer}, \binits{L.}},
\bauthor{\bsnm{Brink}, \binits{T.}},
\bauthor{\bsnm{Freitas}, \binits{R.}},
\bauthor{\bsnm{Frolov}, \binits{T.}},
\bauthor{\bsnm{Dehm}, \binits{G.}},
\bauthor{\bsnm{Liebscher}, \binits{C.H.}}:
\batitle{{Dual phase patterning during a congruent grain boundary phase
  transition in elemental copper}}.
\bjtitle{Nature Communications}
\bvolume{13}(\bissue{1}),
\bfpage{1}--\blpage{11}
(\byear{2022})
{\href{https://arxiv.org/abs/2109.15192}{{arXiv:2109.15192}}}.
\doiurl{10.1038/s41467-022-30922-3}
\end{barticle}
\endbibitem

\bibitem{Ference1988}
\begin{barticle}
\bauthor{\bsnm{Ference}, \binits{T.G.}},
\bauthor{\bsnm{Balluffi}, \binits{R.W.}}:
\batitle{{Observation of a reversible grain boundary faceting transition
  induced by changes of composition}}.
\bjtitle{Scripta Metallurgica}
\bvolume{22}(\bissue{12}),
\bfpage{1929}--\blpage{1934}
(\byear{1988}).
\doiurl{10.1016/S0036-9748(88)80240-0}
\end{barticle}
\endbibitem

\bibitem{Peter2021}
\begin{barticle}
\bauthor{\bsnm{Peter}, \binits{N.J.}},
\bauthor{\bsnm{Duarte}, \binits{M.J.}},
\bauthor{\bsnm{Kirchlechner}, \binits{C.}},
\bauthor{\bsnm{Liebscher}, \binits{C.H.}},
\bauthor{\bsnm{Dehm}, \binits{G.}}:
\batitle{{Faceting diagram for Ag segregation induced nanofaceting at an
  asymmetric Cu tilt grain boundary}}.
\bjtitle{Acta Materialia}
\bvolume{214},
\bfpage{116960}
(\byear{2021}).
\doiurl{10.1016/j.actamat.2021.116960}
\end{barticle}
\endbibitem

\bibitem{Sigle2006}
\begin{barticle}
\bauthor{\bsnm{Sigle}, \binits{W.}},
\bauthor{\bsnm{Richter}, \binits{G.}},
\bauthor{\bsnm{R{\"{u}}hle}, \binits{M.}},
\bauthor{\bsnm{Schmidt}, \binits{S.}}:
\batitle{{Insight into the atomic-scale mechanism of liquid metal
  embrittlement}}.
\bjtitle{Applied Physics Letters}
\bvolume{89}(\bissue{12}),
\bfpage{1}--\blpage{4}
(\byear{2006}).
\doiurl{10.1063/1.2356322}
\end{barticle}
\endbibitem

\bibitem{Zhao2022}
\begin{barticle}
\bauthor{\bsnm{Zhao}, \binits{H.}},
\bauthor{\bsnm{Chakraborty}, \binits{P.}},
\bauthor{\bsnm{Ponge}, \binits{D.}},
\bauthor{\bsnm{Hickel}, \binits{T.}},
\bauthor{\bsnm{Sun}, \binits{B.}},
\bauthor{\bsnm{Wu}, \binits{C.-H.}},
\bauthor{\bsnm{Gault}, \binits{B.}},
\bauthor{\bsnm{Raabe}, \binits{D.}}:
\batitle{Hydrogen trapping and embrittlement in high-strength al alloys}.
\bjtitle{Nature}
\bvolume{602}(\bissue{7897}),
\bfpage{437}--\blpage{441}
(\byear{2022})
\end{barticle}
\endbibitem

\bibitem{Zhang2022}
\begin{barticle}
\bauthor{\bsnm{Zhang}, \binits{S.}},
\bauthor{\bsnm{Xie}, \binits{Z.}},
\bauthor{\bsnm{Keuter}, \binits{P.}},
\bauthor{\bsnm{Ahmad}, \binits{S.}},
\bauthor{\bsnm{Abdellaoui}, \binits{L.}},
\bauthor{\bsnm{Zhou}, \binits{X.}},
\bauthor{\bsnm{Cautaerts}, \binits{N.}},
\bauthor{\bsnm{Breitbach}, \binits{B.}},
\bauthor{\bsnm{Aliramaji}, \binits{S.}},
\bauthor{\bsnm{Korte-Kerzel}, \binits{S.}},
\bauthor{\bsnm{Hans}, \binits{M.}},
\bauthor{\bsnm{Schneider}, \binits{J.M.}},
\bauthor{\bsnm{Scheu}, \binits{C.}}:
\batitle{{Atomistic Structures of 0001 Tilt Grain Boundaries in a Textured Mg
  Thin Film}}.
\bjtitle{Nanoscale}
(\byear{2022}).
\doiurl{10.1039/D2NR05505H}
\end{barticle}
\endbibitem

\bibitem{Wang1997}
\begin{barticle}
\bauthor{\bsnm{Wang}, \binits{Y.C.}},
\bauthor{\bsnm{Ye}, \binits{H.Q.}}:
\batitle{{On the tilt grain boundaries in hcp Ti with [0001] orientation}}.
\bjtitle{Philosophical Magazine A: Physics of Condensed Matter, Structure,
  Defects and Mechanical Properties}
\bvolume{75}(\bissue{1}),
\bfpage{261}--\blpage{272}
(\byear{1997}).
\doiurl{10.1080/01418619708210294}
\end{barticle}
\endbibitem

\bibitem{Ashby1978}
\begin{barticle}
\bauthor{\bsnm{Ashby}, \binits{M.F.}},
\bauthor{\bsnm{Spaepen}, \binits{F.}},
\bauthor{\bsnm{Williams}, \binits{S.}}:
\batitle{The structure of grain boundaries described as a packing of
  polyhedra}.
\bjtitle{Acta Metallurgica}
\bvolume{26}(\bissue{11}),
\bfpage{1647}--\blpage{1663}
(\byear{1978}).
\doiurl{10.1016/0001-6160(78)90075-5}
\end{barticle}
\endbibitem

\bibitem{Pond1979}
\begin{barticle}
\bauthor{\bsnm{Pond}, \binits{R.C.}},
\bauthor{\bsnm{Vitek}, \binits{V.}},
\bauthor{\bsnm{Smith}, \binits{D.A.}}:
\batitle{{Grain boundary structures in f.c.c. and b.c.c. metals and sites for
  segregated impurities}}.
\bjtitle{Acta Crystallographica Section A}
\bvolume{35}(\bissue{4}),
\bfpage{689}--\blpage{693}
(\byear{1979}).
\doiurl{10.1107/S0567739479001571}
\end{barticle}
\endbibitem

\bibitem{Sutton1989}
\begin{barticle}
\bauthor{\bsnm{Sutton}, \binits{A.P.}}:
\batitle{On the structural unit model of grain boundary structure}.
\bjtitle{Philosophical Magazine Letters}
\bvolume{59}(\bissue{2}),
\bfpage{53}--\blpage{59}
(\byear{1989}).
\doiurl{10.1080/09500838908214777}
\end{barticle}
\endbibitem

\bibitem{huber_Mg}
\begin{barticle}
\bauthor{\bsnm{Huber}, \binits{L.}},
\bauthor{\bsnm{Rottler}, \binits{J.}},
\bauthor{\bsnm{Militzer}, \binits{M.}}:
\batitle{Atomistic simulations of the interaction of alloying elements with
  grain boundaries in mg}.
\bjtitle{Acta materialia}
\bvolume{80},
\bfpage{194}--\blpage{204}
(\byear{2014})
\end{barticle}
\endbibitem

\bibitem{Sato2007}
\begin{barticle}
\bauthor{\bsnm{Sato}, \binits{Y.}},
\bauthor{\bsnm{Yamamoto}, \binits{T.}},
\bauthor{\bsnm{Ikuhara}, \binits{Y.}}:
\batitle{{Atomic structures and electrical properties of ZnO grain
  boundaries}}.
\bjtitle{Journal of the American Ceramic Society}
\bvolume{90}(\bissue{2}),
\bfpage{337}--\blpage{357}
(\byear{2007}).
\doiurl{10.1111/j.1551-2916.2006.01481.x}
\end{barticle}
\endbibitem

\bibitem{Feng2009}
\begin{barticle}
\bauthor{\bsnm{Feng}, \binits{Y.}},
\bauthor{\bsnm{Wang}, \binits{R.}},
\bauthor{\bsnm{Liu}, \binits{H.}},
\bauthor{\bsnm{Jin}, \binits{Z.}}:
\batitle{{Thermodynamic reassessment of the magnesium-gallium system}}.
\bjtitle{Journal of Alloys and Compounds}
\bvolume{486}(\bissue{1-2}),
\bfpage{581}--\blpage{585}
(\byear{2009}).
\doiurl{10.1016/j.jallcom.2009.07.010}
\end{barticle}
\endbibitem

\bibitem{Zhang2018}
\begin{barticle}
\bauthor{\bsnm{Zhang}, \binits{S.}},
\bauthor{\bsnm{Scheu}, \binits{C.}}:
\batitle{{Evaluation of EELS spectrum imaging data by spectral components and
  factors from multivariate analysis}}.
\bjtitle{Microscopy}
\bvolume{67},
\bfpage{133}--\blpage{141}
(\byear{2018}).
\doiurl{10.1093/jmicro/dfx091}
\end{barticle}
\endbibitem

\bibitem{vasp_1}
\begin{barticle}
\bauthor{\bsnm{Kresse}, \binits{G.}},
\bauthor{\bsnm{Hafner}, \binits{J.}}:
\batitle{Ab initio molecular dynamics for liquid metals}.
\bjtitle{Phys. Rev. B}
\bvolume{47},
\bfpage{558}--\blpage{561}
(\byear{1993}).
\doiurl{10.1103/PhysRevB.47.558}
\end{barticle}
\endbibitem

\bibitem{vasp_2}
\begin{barticle}
\bauthor{\bsnm{Kresse}, \binits{G.}},
\bauthor{\bsnm{Furthm{\"u}ller}, \binits{J.}}:
\batitle{Efficient iterative schemes for ab initio total-energy calculations
  using a plane-wave basis set}.
\bjtitle{Physical review B}
\bvolume{54}(\bissue{16}),
\bfpage{11169}
(\byear{1996})
\end{barticle}
\endbibitem

\bibitem{vasp_3}
\begin{barticle}
\bauthor{\bsnm{Kresse}, \binits{G.}},
\bauthor{\bsnm{Joubert}, \binits{D.}}:
\batitle{From ultrasoft pseudopotentials to the projector augmented-wave
  method}.
\bjtitle{Physical review b}
\bvolume{59}(\bissue{3}),
\bfpage{1758}
(\byear{1999})
\end{barticle}
\endbibitem

\bibitem{monkhorst_pack}
\begin{barticle}
\bauthor{\bsnm{Monkhorst}, \binits{H.J.}},
\bauthor{\bsnm{Pack}, \binits{J.D.}}:
\batitle{Special points for brillouin-zone integrations}.
\bjtitle{Physical review B}
\bvolume{13}(\bissue{12}),
\bfpage{5188}
(\byear{1976})
\end{barticle}
\endbibitem

\bibitem{methfessel_paxton}
\begin{barticle}
\bauthor{\bsnm{Methfessel}, \binits{M.}},
\bauthor{\bsnm{Paxton}, \binits{A.T.}}:
\batitle{High-precision sampling for brillouin-zone integration in metals}.
\bjtitle{Phys. Rev. B}
\bvolume{40},
\bfpage{3616}--\blpage{3621}
(\byear{1989}).
\doiurl{10.1103/PhysRevB.40.3616}
\end{barticle}
\endbibitem

\bibitem{Alhasan2023}
\begin{barticle}
\bauthor{\bsnm{Alhasan}, \binits{A.S.A.}},
\bauthor{\bsnm{Zhang}, \binits{S.}},
\bauthor{\bsnm{Berkels}, \binits{B.}}:
\batitle{Direct motif extraction from high resolution crystalline {STEM}
  images}.
\bjtitle{Ultramicroscopy}
\bvolume{254},
\bfpage{113827}
(\byear{2023})
{\href{https://arxiv.org/abs/2303.07438}{{arXiv:2303.07438}}}
{[eess.IV]}.
\doiurl{10.1016/j.ultramic.2023.113827}
\end{barticle}
\endbibitem

\bibitem{Allen2015}
\begin{barticle}
\bauthor{\bsnm{Allen}, \binits{L.J.}},
\bauthor{\bsnm{D'Alfonso}, \binits{A.J.}},
\bauthor{\bsnm{Findlay}, \binits{S.D.}}:
\batitle{Modelling the inelastic scattering of fast electrons}.
\bjtitle{Ultramicroscopy}
\bvolume{151},
\bfpage{11}--\blpage{22}
(\byear{2015}).
\doiurl{10.1016/j.ultramic.2014.10.011}.
\bcomment{Special Issue: 80th Birthday of Harald Rose; PICO 2015 – Third
  Conference on Frontiers of Aberration Corrected Electron Microscopy}
\end{barticle}
\endbibitem

\end{thebibliography}

\end{document}